# Legacy Learning Strategy Based on Few-Shot Font Generation Models for Automatic Text Design in Metaverse Content

Younghwi Kim[a], Dohee Kim[b], Seok Chan Jeong[c], Sunghyun Sim[a,d †]
*[a] Graduate School of AI Convergence Engineering, Changwon National University*
*20 Changwondaehak-ro Uichang Gu, 51140, Changwon, South Korea*
*[b]Safe & Clean Supply Chain Research Center, Pusan National University,*
*30-Jan-jeon Dong, Geum-Jeong Gu, 46241, Busan, South Korea*
*[c]Department of e-Business & AI Grand ICT Research Center, Dong-eui University,*
*176 Eomgwang No, Gaya Dong 24, Busanjin Gu, 47340 Busan, South Korea*
*[d] School of AI Convergence Engineering, Changwon National University*
*20 Changwondaehak-ro Uichang Gu, 51140, Changwon, South Korea*
*† Corresponding Author*

**Abstract.** The metaverse consists of hardware, software, and content, among which text design plays a critical role in enhancing user immersion and usability as a content element. However, in languages such as Korean and Chinese, which require thousands of unique glyphs, creating new text designs involves high costs and complexity. To address this, this study proposes a training strategy called legacy learning, which recombines and transforms structures based on existing text design models. This approach enables the generation of new text designs and improves quality without manual design processes. To evaluate legacy learning, it was applied to Korean and Chinese text designs. Additionally, we compared the results before and after seven state-of-the-art text-generation models. Consequently, text designs generated via legacy learning resulted in more than a 30% difference in Fréchet Inception Distance (FID) and Learned Perceptual Image Patch Similarity (LPIPS) metrics compared with the original methods and presented meaningful style variations in visual comparisons. Furthermore, the repeated learning process improved the structural consistency of the generated characters, and an OCR-based evaluation showed increasing recognition accuracy across iterations, indicating improved legibility of the generated glyphs. In addition, a system usability scale (SUS) survey was conducted to evaluate usability among metaverse content designers and general users. The expert group recorded a score of 95.78 ("Best Imaginable"), whereas the nonexpert group scored 76.42 ("Excellent"), indicating an overall high level of usability. These results suggest that legacy learning can significantly improve both the production efficiency and quality of text design in the metaverse environment.

## 1. Introduction

The metaverse is an immersive digital space that combines virtual reality (VR) with the real world and is powered by digital technologies (Dwivedi et al., 2022). It enables people to engage in various social activities, such as gaming, watching movies, specialized education, remote collaboration, and marketing (Mancuso et al., 2023; Kim et al., 2025; Widodo et al., 2024). The components of a metaverse can be divided into three main aspects: hardware, software, and content (Park & Kim, 2022). Among these, content consists of five elements: 1) virtual environments, 2) avatars and interfaces, 3) social interactions and communities, 4) storytelling, and 5) linguistic elements and text design. Linguistic elements, particularly text design, play a central role in shaping user experience and immersion. They enable the representation of identity, emotional tone, and cultural nuance in virtual spaces.

However, text design in the metaverse is highly dependent on the language being used, and the complexity of script systems leads to varying levels of workload for designers. For example, in English, which consists of 26 characters, 26 lowercase letters and 26 uppercase letters are sufficient for completing text design requirements. By contrast, Korean requires designing combinations of 19 onsets, 21 nuclei, and 28 codons, resulting in 11,172 possible character combinations (19 × 21 × 28). Figure 1 illustrates how Chinese and Korean characters are constructed, including component labels. To design the Korean character “값”, the designer must first create its components: the onset “ㄱ”, the nucleus “ㅏ”, and the coda “ㅄ”. These elements are then combined to form the final glyph “값”, which is assigned a specific text design. Chinese characters present an even greater challenge. Unlike compositional Korean characters, Chinese glyphs follow a radical-based structure, but with more than 60,000 characters officially defined, font design becomes overwhelmingly labor intensive. Designers are required to construct each glyph individually, making traditional manual font design for such languages highly inefficient. This linguistic complexity underscores the need for automated, scalable, and adaptive font generation strategies to support immersive multilingual content in the metaverse.

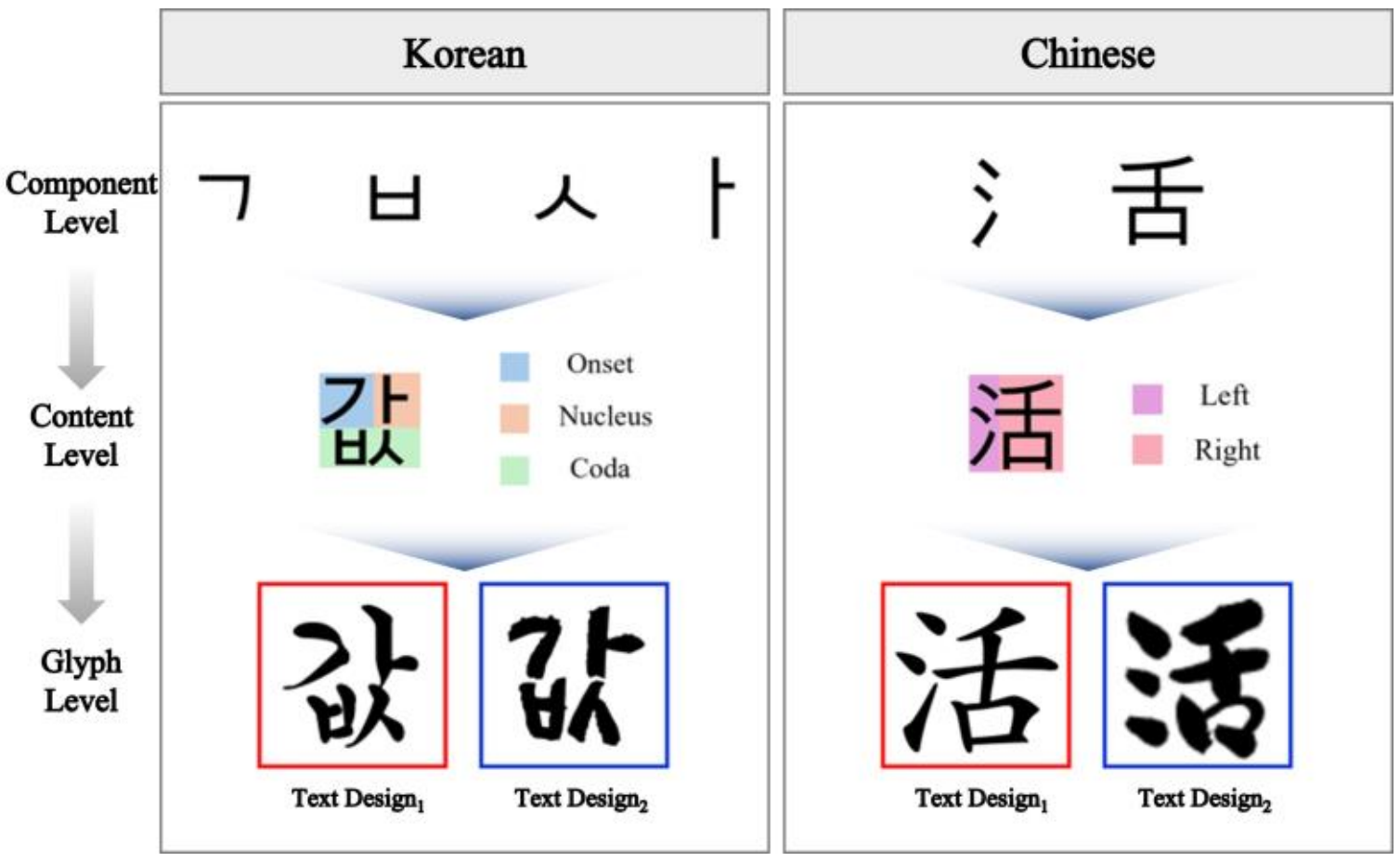

**Figure 1**. Example of the text design process for Korean and Chinese.

These issues can inevitably arise not only in metaverse environments but also during the development of various online content, games, and VR/augmented reality (AR) applications. To address these issues, recent research has actively explored few-shot font generation (FFG) models via generative AI. This approach requires designers to manually create only a minimal subset of glyphs, while the remaining glyphs are automatically synthesized by the model. This paradigm represents a shift away from traditional methods that require the design of all 11,172 Korean or 60,000 Chinese glyphs per font. Instead, FFG enables the design of only 5–10% of the glyphs, significantly improving design efficiency (Chen et al., 2024; Lee & Choi, 2024). However, despite its efficiency, FFG has practical limitations. As the number of input glyphs decreases, the quality of the generated fonts tends to degrade. Studies suggest that to maintain legibility and consistency, at least 10% of the full glyph set (approximately 1,100 for Korean and 6,000 for Chinese) is required for training. This high input requirement remains a significant burden for designers, who must create multiple text designs across different languages for multilingual metaverse content, thereby posing a major bottleneck to the practical deployment of FFG models.

To address these limitations, this study proposes a plug-and-play legacy learning strategy designed to improve the performance of existing FFG models without modifying their architecture. Rather than designing a new model from scratch, this strategy integrates a repetitive training-based transfer mechanism into existing FFG systems. By leveraging the natural stylistic variation that occurs between inputs and outputs in generative models, the method applies controlled transformations across multiple generations to gradually increase both the diversity and quality of text design.

Specifically, the legacy learning strategy first trains a model using a complete glyph set. When the generated outputs exhibit slight stylistic differences from the target text design, these outputs are then used as the input for the next generation. This process is repeated over N generations while maintaining full glyph-based refinement throughout. Consequently, the structural consistency of characters is preserved, while visual diversity progressively increases. The proposed framework is generic and modular, allowing easy integration with various languages and FFG backbones, and is well suited for scalable deployment in metaverse content-creation environments. The main contributions of this study are as follows:

- **(C1)** We propose a plug-and-play legacy learning strategy that can be integrated into existing FFG models without any structural modifications, thereby enabling automated text design for multilingual metaverse content.
- **(C2)** Through quantitative experiments, we confirm that legacy learning generates text designs that are stylistically distinct from the original designs and that it improves character-level quality as well.
- **(C3)** Visual analysis confirms that the glyphs generated through legacy learning consistently maintain the global layout and stroke-level quality, even as their stylistic features evolve.
- **(C4)** A system usability scale (SUS) study involving 35 professional designers and 35 non-expert users yielded average scores of 95.78 and 76.42, respectively, confirming the usability and applicability of the proposed system in real-world metaverse text design tasks.

The remainder of this paper is as follows. Sections 2 and 3 review related work and define the objectives of this study. Section 4 introduces the concept of the proposed method, Legacy Learning, and presents the design results of the user interface and user experience (UX) based on this method. Section 5 provides a detailed description of the experimental settings and datasets used, whereas Section 6 presents both quantitative and qualitative experimental results and discusses the practical applicability of the method, including the results of an SUS test. Finally, Section 7 discusses the theoretical and practical implications of this study, and Section 8 summarizes the key findings and proposes directions for future research.

## 2. Related work

This section covers related work on components of a metaverse, text design in 3D virtual environments, such as metaverses, and related work on generative AI models for text design in virtual spaces.

## 2.1. Components of a Metaverse

The metaverse consists of three primary components: hardware, software, and content (Park & Kim, 2022). Each element has distinct characteristics and roles, working synergistically to enhance the functionality and user experience of the metaverse (Kim et al., 2025). The hardware component includes various devices that enable users to perceive and interact with virtual environments (Richter & Richter, 2023). Key devices include headsets that provide extended reality (XR) experiences, such as VR, AR, and mixed reality. These devices are essential for building immersive environments (Xi et al., 2023; Kim et al., 2022; Pacchierotti et al., 2024). On the software side, a range of technologies is used to implement and optimize metaverse experiences (Park & Kim, 2022; Cappannari & Vitillo, 2022; Wang et al., 2022). Procedural content generation techniques allow the efficient creation of high-quality three-dimensional (3D) content, significantly reducing the time and cost required to construct large-scale virtual environments (Mourtzis et al., 2022). Additionally, AI-based systems play a crucial role in managing XR applications, particularly in complex multi-user environments, ensuring that the user experience is seamless and engaging (Huynh-The et al., 2023).

Content encompasses diverse activities and services that users can experience within a metaverse (Zallio & Clarkson, 2022; Park & Kim, 2022). The detailed elements of metaverse content include the following: 1) ***Virtual environments***: worlds that users can explore, offering varied landscapes and interactive elements (Sung et al., 2021; Dwivedi et al., 2022; Richter & Richter, 2023); 2) ***Avatars and interfaces***: tools that enable interaction between users and the virtual world, representing users in the digital space (Li & Liu, 2019; Wu, 2015); and 3) ***Social interactions and communities***: platforms for users to engage in social experiences, including communication, collaboration, and community building (López-Cabarcos & Piñeiro-Chousa, 2024; Sayari et al., 2025; Zhang & Ye, 2023). 4) ***Storytelling***: narratives and scenarios that enable users to deeply immerse themselves in a virtual environment, enhancing their emotional responses and engagement while making their experiences within the virtual world more meaningful and personal (Choi & Kim, 2017; Mogaji et al., 2023; Ruusunen et al., 2023). 5) ***Linguistic elements and text design***: essential components that facilitate interaction and communication within the virtual world, contributing to an overall immersive experience (Chen, 2022; Sun et al., 2023; Morales-Fernández, 2024). Among these, text design is a crucial yet often underexplored element. It supports clear communication, improves readability, and reinforces the immersive qualities of metaverse environments (Yang, 2021; Xu, 2022; Elhagry, 2023; Woodward & Ruiz, 2023; He et al., 2024a).

## 2.2. Text design in 3D virtual environments

In 3D virtual environments such as VR, AR, and metaverse platforms, text design plays a significant role in enhancing readability, task performance, and user engagement (Kojic et al., 2022; Seinfeld et al., 2020; Shi et al., 2020; De Back et al., 2023).

Knaack et al. (2019) demonstrated that text design within a VR interface affects readability and directly influences the user response time. Similarly, Wang et al. (2020) demonstrated that text design preferences vary among users and that applying text designs that match individual preferences can enhance both readability and immersion. Wei et al. (2020) examined how text design and other interface elements, such as voice, in 3D virtual environments, such as metaverse environments, affect user immersion, communication abilities, and satisfaction. This study demonstrates that text design directly and indirectly affects immersion, readability, and communication abilities.

As demonstrated in various related studies, text design in 3D virtual environments is key in enhancing user readability, communication capabilities, and content immersion. Although designers often rely on commercial fonts or subscriptions (Elhagry, 2023; Hartini & Awaliyah, 2023), text design preferences vary significantly among users (Dudley et al., 2023), and mass-produced fonts rarely satisfy all users' needs. Thus, generating personalized fonts that support real-time immersion and communication in 3D space has become a research priority.

### 2.3. Few-shot font generation models

As introduced in Section 2.2, research based on generative AI is being actively conducted to address the issues of high cost and time consumption associated with text design. Generative AI models for text design are classified into MFG and FFG models, both of which utilize image-to-image (I2I) translation methods (Park et al., 2021a; Tang et al., 2022; Liu et al., 2022). I2I translation is a visual task focused on mapping images from one domain to another while preserving the image content (Liu et al., 2022). In general, I2I translation has been applied in various fields, such as image animation, object transfiguration, and semantic segmentation, and is also actively used in text design generation (Xie et al., 2021; Tang et al., 2022). In text design generation, I2I translation methods are used to learn the shape of specific text designs and convert the input glyphs into learned text designs while preserving text content. This approach was employed in both the MFG and FFG models.

FFG models were introduced to overcome MFG's cost barrier by learning from a limited number of glyphs (Cha et al., 2020; Park et al., 2021a). Early models such as the encoder-mixer-decoder and artistic glyph image synthesis network (AGISNet) extracted glyph-level features (Zhang et al., 2018; Gao et al., 2019), whereas more recent models such as DM-Font, LF-Font, and MX-Font added context-aware representations with each learning feature at the component, contextual, and glyph levels, respectively (Cha et al., 2020; Park et al., 2021a; Park et al., 2021b). CF-Font employs feature deformation skip connections to predict a displacement map from the content glyph to the target style glyph and uses the CFM module to fuse multiple content features (Wang et al., 2023). Diff-Font applies a diffusion model to font generation, demonstrating accurate generation of complex characters through a conditional generation task that extracts font style via predefined embedding tokens and one-shot learning (He et al., 2024b). VQ-Font mitigates issues of missing details and distorted strokes in character generation by refining font images through the integration of token prioritization encapsulated in a pretrained font codebook (Yao et al., 2024). This advancement enables the generation of complete text designs from a small number of glyphs, even for languages with a large number of text content, such as Korean and Chinese. Additionally, FontDiffuser is an image-to-image font generation method based on a diffusion model that models the font imitation task as a noise-to-denoise paradigm. This approach introduces a multiscale content aggregation (MCA) block, which effectively integrates global and local content cues across various scales, enabling enhanced preservation of intricate strokes in complex characters (Yang et al., 2024).

Despite these achievements, FFG models still require at least 5–10% of the total glyph set for languages such as Korean or Chinese, amounting to thousands of glyphs (Hassan et al., 2024; He et al., 2024c). This requirement imposes a practical bottleneck on real-world applications where user-specific fonts or dynamic updates are needed. To address these limitations, this study proposes a new text design learning strategy called Legacy Learning that can directly generate high-quality, user-specific text designs and offers a text design autogeneration service for metaverse content.

## 3. Research objective

As discussed previously, Figure 2 intuitively illustrates these limitations. Regardless of the model used to generate the text designs, fewer glyphs used in training tended to result in more noise in the generated text designs. In contrast, increasing the number of glyphs improves the output quality. The visualization results indicate that while training on many glyphs can produce high-quality designs, achieving acceptable levels of noise reduction and design fidelity typically requires training on at least 10% of the glyph set (Hassan et al., 2024; He et al., 2024c). For Korean and Chinese individuals, this number corresponds to approximately 1,100 and 6,000 glyphs, respectively. Such requirements present a major issue for scalable, automated text design generation for metaverse applications.

| Model Type | Num. of Train Glyph | Train Glyph | Generation Glyph Results | Comparison of the Quality of Glyph Generation Results |
|---|---|---|---|---|
| MX-Font | 18 | | | |
| MX-Font | 72 | | | |
| MX-Font | 108 | | | |
| DM-Font | 18 | | | |
| DM-Font | 72 | | | |
| DM-Font | 108 | | | |
| LF-Font | 18 | | | |
| LF-Font | 72 | | | |
| LF-Font | 108 | | | |

**Figure 2**. Visualization results of generated font designs based on the increase in glyphs used for training.

On the basis of these issues, we aim to address the following research questions:

- **(RQ1)** Can a new learning strategy, unlike traditional FFG models that rely on designer-provided partial glyph sets that transform existing completed text designs to generate new text designs?
- **(RQ2)** Can the legacy learning-based text design generation method proposed to address RQ1 produce text designs of higher quality than those generated by conventional FFG models?
- **(RQ3)** Can the automated system based on the proposed legacy learning strategy be practically applied to metaverse content creation in a meaningful way?

## 4. Legacy Learning for automatic text design generation in metaverse environments

In this section, we introduce the legacy learning strategy for training text-design generation models in metaverse content and a service that can provide personalized text designs for each user via this training model. Table 1 summarizes the notation used in legacy learning.

**Table 1**. Summary table of the notations used in legacy learning.

| Notation | Description | Notation | Description |
|---|---|---|---|
| $s$ | Style of source glyphs | $f$ | Extracted feature |
| $\hat{s}$ | Style of reference glyphs | $x_{\hat{s},C}$ | Reference full glyph set |
| $C$ | Full content | $\hat{x}_{\hat{s},C}$ | Generated full glyph set |
| $c$ | Few-shot content | $x_{\hat{s},c}$ | Reference few-shot glyph set |

## 4.1. Overview of the FFG network for Legacy Learning

The legacy learning strategy employs FFG as its primary concept. The FFG approach typically adopts a learning method based on generative adversarial networks. In this method, two components are adversarially trained: an encoder and decoder generator (***G***) that generates images with a distribution similar to the target image and a discriminator (***D)*** and a component classifier (***CLS***) that assist in training ***G***. When ***G*** generates glyphs similar to the reference glyphs (ground truth), ***D*** and ***CLS*** evaluate the quality of the generated glyphs by comparing it with the reference glyphs. This feedback helps ***G*** ultimately generate images that closely resemble target images. This approach allows ***G*** to progressively learn and generate images with distributions that are increasingly similar to reality.

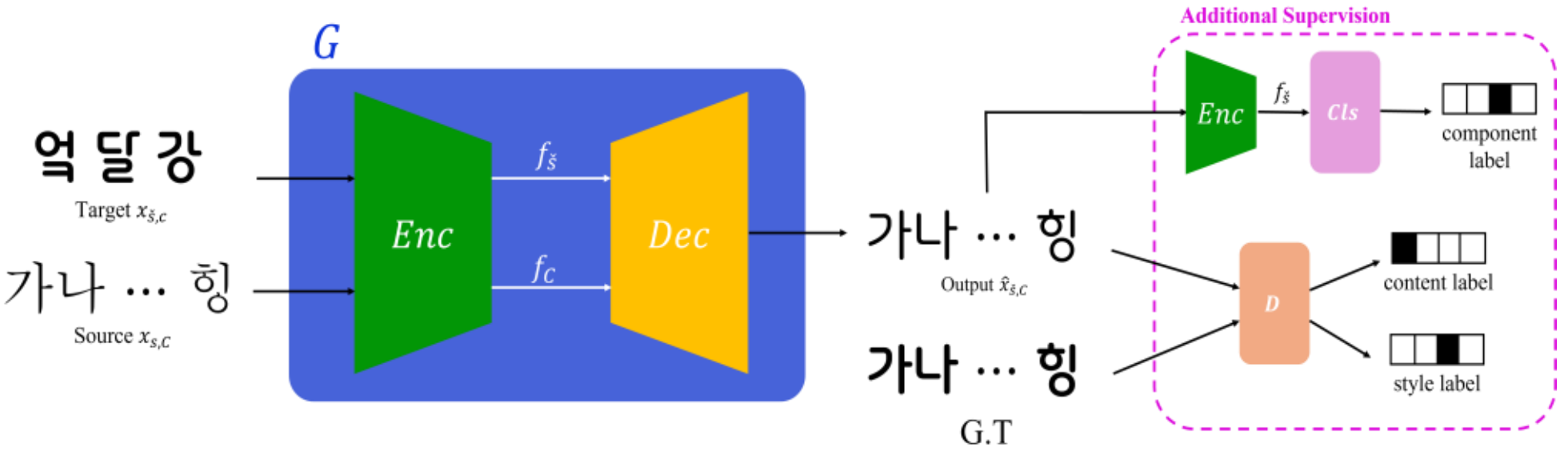

**Figure 3.** Structure of the FFG network.

Figure 3 illustrates the process described above. For example, when learning a style using three few-shot reference glyphs, ***G*** attempts to generate all glyphs by applying the learned style to the content. Each content is extracted from a full content set provided by the source, which is typically fixed as a default font. However, ***G*** may generate styles or content that differ from the target. At this point, ***D*** and ***CLS*** provide additional supervision for ***G*** by evaluating the quality of the generated glyphs.

### 4.1.1. Generator design

Generator ***G*** consists of an encoder ***Enc*** and a decoder ***Dec***. ***Enc*** consists of ***Enc$_c$***, which learns the content features of the text, and ***Enc$_s$***, which learns the style features of the text design. ***Enc$_c$*** aims to extract the content feature $f_C$ for each glyph from the full-glyph set of the source (default), whereas ***Enc$_s$*** focuses on extracting style features from the reference few-shot glyph set. The features $f_{\hat{s}}$ and $f_C$ extracted through this process are used by ***Dec*** to generate new text designs. This process can be represented as follows:

$$f_{\hat{s}} = Enc_s\left(x_{\hat{s},c}\right) \text{ and } f_C = Enc_c\left(x_{s,C}\right), \tag{1}$$

$$\hat{x}_{\hat{s},C} = Dec(f_{\hat{s}}, f_C), \tag{2}$$

$$G\left(x_{\hat{s},c}, x_{s,C}\right) = Dec\left(Enc_s\left(x_{\hat{s},c}\right), Enc_C\left(x_{s,C}\right)\right) = \hat{x}_{\hat{s},C}. \tag{3}$$

### 4.1.2. Discriminator design

Discriminator ***D*** uses a multitask approach that assesses both style and content, enhancing the quality of the outputs generated by ***G***. During training, the inputs are $x_{\hat{s},C}$, which denotes the full-glyph set of the reference text design and the generated $\hat{x}_{\hat{s},C}$. These inputs are classified into style and content labels. When $\hat{x}_{\hat{s},C}$ is the input, it is evaluated as $D(\hat{x}_{\hat{s},C})$, whereas $\hat{x}_{\hat{s},C}$ is evaluated as $D(x_{\hat{s},C})$. This process provides feedback to ***G*** throughout the training.

#### 4.1.3. Component classifier design

A ***CLS*** provides additional supervision for ***G*** by performing component classification for glyphs with local components. During training, the features of the glyph are extracted from the generated output $\hat{x}_{\hat{s},C}$ via the trained ***Enc***. These features are classified and compared with the actual components of $x_{\hat{s},C}$. This is represented by $CLS(f_{\hat{s},\hat{x}})$ and $CLS(f_{\hat{s},x})$.

### 4.2. Iterative Legacy Learning strategy for style-enhanced glyph generation

The FFG method involves learning the general style patterns of the reference text design used during training and applying these learned styles to the remaining content. In this process, the generated glyphs may exhibit slight stylistic differences from the reference glyphs. Leveraging this characteristic of FFG, we propose the legacy learning approach, which enables the generation of high-quality text designs that differ from the original designs.

Figure 4 illustrates that the training structure of legacy learning consists of the following steps. ***Step I***: Construct a dataset comprising diverse combinations of text designs and train an FFG model. In this stage, 60% of the glyphs from each reference text design are used as inputs for style extraction, whereas the remaining 40% are used as ground truths to supervise the model by comparing them with the generated glyphs. This enables supervised learning to generate the full glyph set. ***Step II***: Use the initially trained legacy FFG model to generate all glyphs from the reference text design used in training, thereby constructing a new text design set. ***Step III***: Perform transfer learning on the legacy FFG model using the generated text designs from the previous step as new training data. ***Step IV***: Repeat Steps II and III until the design variations across the entire text design set converge. The number of repetitions is defined as ***N***.

The legacy learning strategy incorporates a generalization process in each iterative training step based on a limited dataset. In the initial stages, the model generates outputs that closely resemble the reference glyphs. However, as the iterations progress, transfer learning based on the outputs of the previous step induces increasing stylistic diversity and visual variation. Furthermore, because the model learns from existing text designs, there is no constraint on the number of input glyphs, allowing the generation of structurally stable text designs.

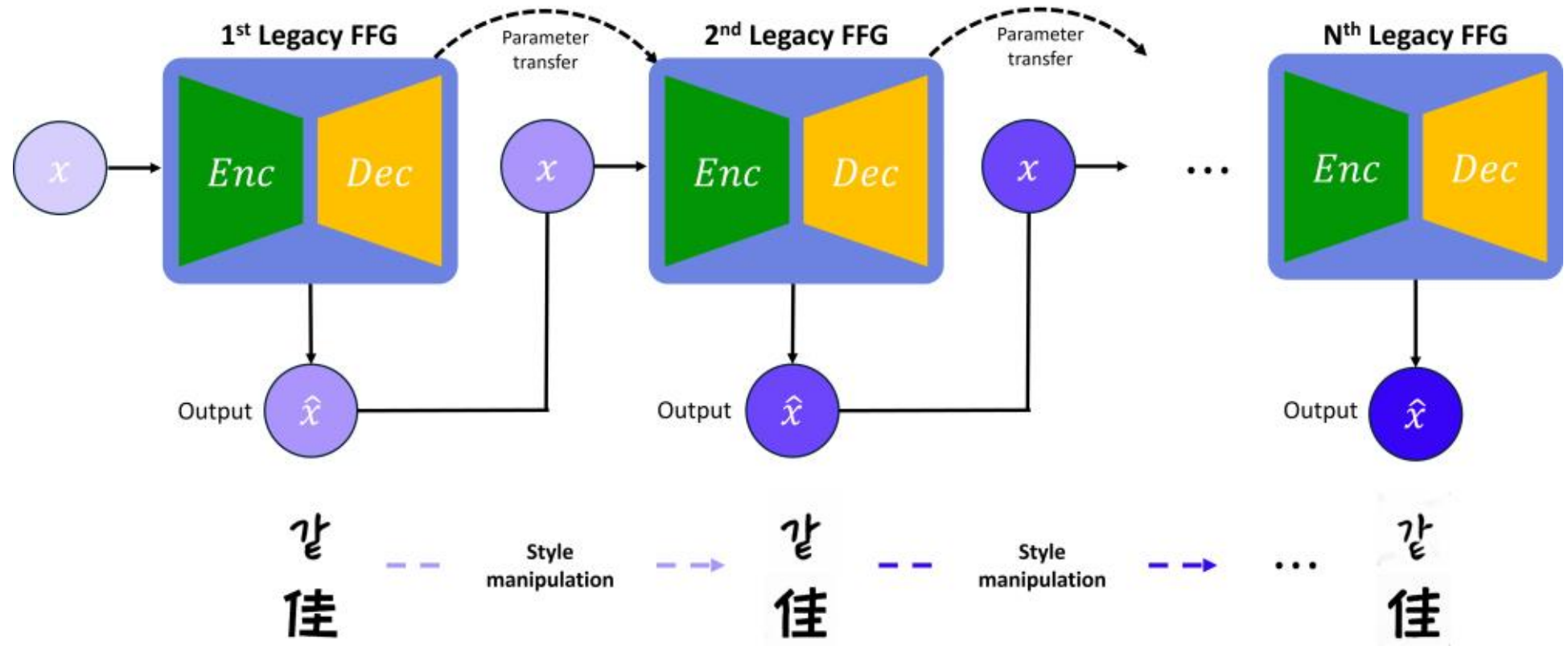

**Figure 4.** Visualization of the overall training structure of Legacy Learning.

#### 4.2.1. Loss functions

In each FFG training, loss functions are applied to train ***G***, ***D***, and ***CLS***, and these loss functions consist of the following four individual loss functions:

***L₁ Loss***. The difference between the actual and generated images is defined as a loss function to minimize the difference between the image generated by $G$ and the actual image. The $L_1$ loss is structured as follows:

$$\mathcal{L}_{l1} = \mathbb{E}_{x,\hat{x}}\left[\left\|x_{\hat{s},C} - \hat{x}_{\hat{s},C}\right\|_1\right]. \tag{4}$$

***Adversarial loss***. The log probability of the image generated by $G$ and the actual image is defined as the loss function. This encourages $G$ to generate more plausible images and enables $D$ to better distinguish between generated and actual images. The adversarial loss is structured as follows:

$$\mathcal{L}_{adv} = \mathbb{E}_{x,\hat{x}}\left[logD\left(x_{\hat{s},C}\right)\right] + \mathbb{E}_{x,\hat{x}}[\log\left(1 - D\left(\hat{x}_{\hat{s},C}\right)\right). \tag{5}$$

***Feature-matching loss***. The feature-matching loss function was defined by calculating the difference between the features of the image generated by ***G*** and the actual image in each layer of ***D***. This aims to minimize the difference in features between the generated and actual images when ***G*** generates the images. The feature-matching loss is structured as follows:

$$\mathcal{L}_{fm} = \mathbb{E}_{x,\hat{x}}\left[\frac{1}{L}\sum_{i=1}^{L}\left\|D^{(i)}\left(x_{\hat{s},C}\right) - D^{(i)}\left(\hat{x}_{\hat{s},C}\right)\right\|_1\right]. \tag{6}$$

***Component classification loss***. The cross-entropy (CE) loss function is defined for $CLS$ to perform multiclass classification of component $u$ from images generated by $G$ and real images. This is intended to help the model generate a clearer composition of the components constituting the glyphs. The component-classification loss is structured as follows. Here, $f_{\hat{s},x}$ and $f_{\hat{s},\hat{x}}$ represent the text design features of the actual image and the image generated by ***G***, respectively; $u$ denotes each component; and $U$ denotes the set of all components.

$$\mathcal{L}_c = \mathbb{E}_{x,\hat{x}}\left[\sum_{u\in U} CE\left(CLS\left(f_{\hat{s},x}\right), u\right]\right. + \mathbb{E}_{\hat{x}}\left[\sum_{u\in U} CE\left(CLS\left(f_{\hat{s},\hat{x}}\right), u\right)\right]. \tag{7}$$

***Full objective***. The final objective function is formulated to optimize ***G***, ***D***, and ***CLS*** via the various types of losses defined above. It is expressed as follows, where $L_G$, $L_D$, and $L_c$ denote the losses for $\boldsymbol{G}$, $\boldsymbol{D}$, and $\boldsymbol{CLS}$, respectively, and $\lambda$ represents the parameters controlling the importance of the loss functions. During the training process, $\boldsymbol{D}$ maximizes its loss, whereas $\boldsymbol{G}$ and $\boldsymbol{CLS}$ minimize it. Furthermore, $\boldsymbol{G}$, $\boldsymbol{D}$, and $\boldsymbol{CLS}$ operate independently during parameter updates on the basis of their respective losses.

$$\mathcal{L}_D = \lambda_{adv}\,\mathcal{L}_{adv},\ \mathcal{L}_{Cls} = \lambda_c\,\mathcal{L}_c, \tag{8}$$

$$\mathcal{L}_G = \lambda_{l1}\,\mathcal{L}_{l1} + \lambda_{adv}\,\mathcal{L}_{adv} + \lambda_{fm}\,\mathcal{L}_{fm}. \tag{9}$$

**Algorithm 1.** Legacy Learning

**Input:** source glyph set $x_{s,C}$, reference glyph set $x_{\hat{s},C}$
**Output:** output set $\hat{x}_{set}$ of the Legacy Learning from the 1st to the $N$th

1: $\hat{x}_{set} = []$
2: **for** $i \leftarrow 1$ to $N$ **do**
3: **for** each epoch **do**
4: $f_C = Enc_c^i(x_{s,C})$ *//extract content feature from* $x_{s,C}$
5: $f_{\hat{s}} = Enc_s^i(x_{\hat{s},c})$ *//extract style feature from* $x_{\hat{s},c}$
6: $\hat{x}_{\hat{s},C} = Dec^i(f_{\hat{s}}, f_C)$ *//generate output image by FFGDec*
7: $\theta_D = \theta_D - \eta_D \nabla_{\theta_D} \mathcal{L}_D$ *//update* $\theta_D$ *by taking optimizer on loss* $\mathcal{L}_D$ *defined eq.(8)*
8: $\theta_{Cls} = \theta_{Cls} - \eta_{Cls} \nabla_{\theta_{Cls}} \mathcal{L}_{Cls}$ *//update* $\theta_{CLS}$ *by taking optimizer on loss* $\mathcal{L}_{CLS}$ *defined eq.(8)*
9: $\theta_G = \theta_G - \eta_G \nabla_{\theta_G} \mathcal{L}_G$ *//update* $\theta_G$ *by taking optimizer on loss* $\mathcal{L}_G$ *defined eq.(9)*
10 **end for**
11: $\hat{x}_{\hat{s},C} = G^i(x_{\hat{s},c}, x_{s,C})$ *//full glyph generation*
12: Append $\hat{x}_{\hat{s},C}$ to $\hat{x}_{set}$
13: $x_{\hat{s},C} \leftarrow \hat{x}_{\hat{s},C}$ *//Update* $\hat{x}_{\hat{s},C}$ *as a new reference*
14: **end for**
15: **return** $\hat{x}_{set}$

Accordingly, the complete legacy learning procedure based on such an FFG model is presented in Algorithm 1. In Algorithm 1, $\theta$, $\eta$, and $\nabla_\theta \mathcal{L}$ represent the parameters, learning rate, and gradient of the loss function for ***D***, ***CLS***, and ***G***, respectively.

## 4.3. Automatic text design generation service for metaverse contents

In this section, we introduce a service framework that leverages the previously discussed Legacy Learning to automatically generate and modify text designs within metaverse content.

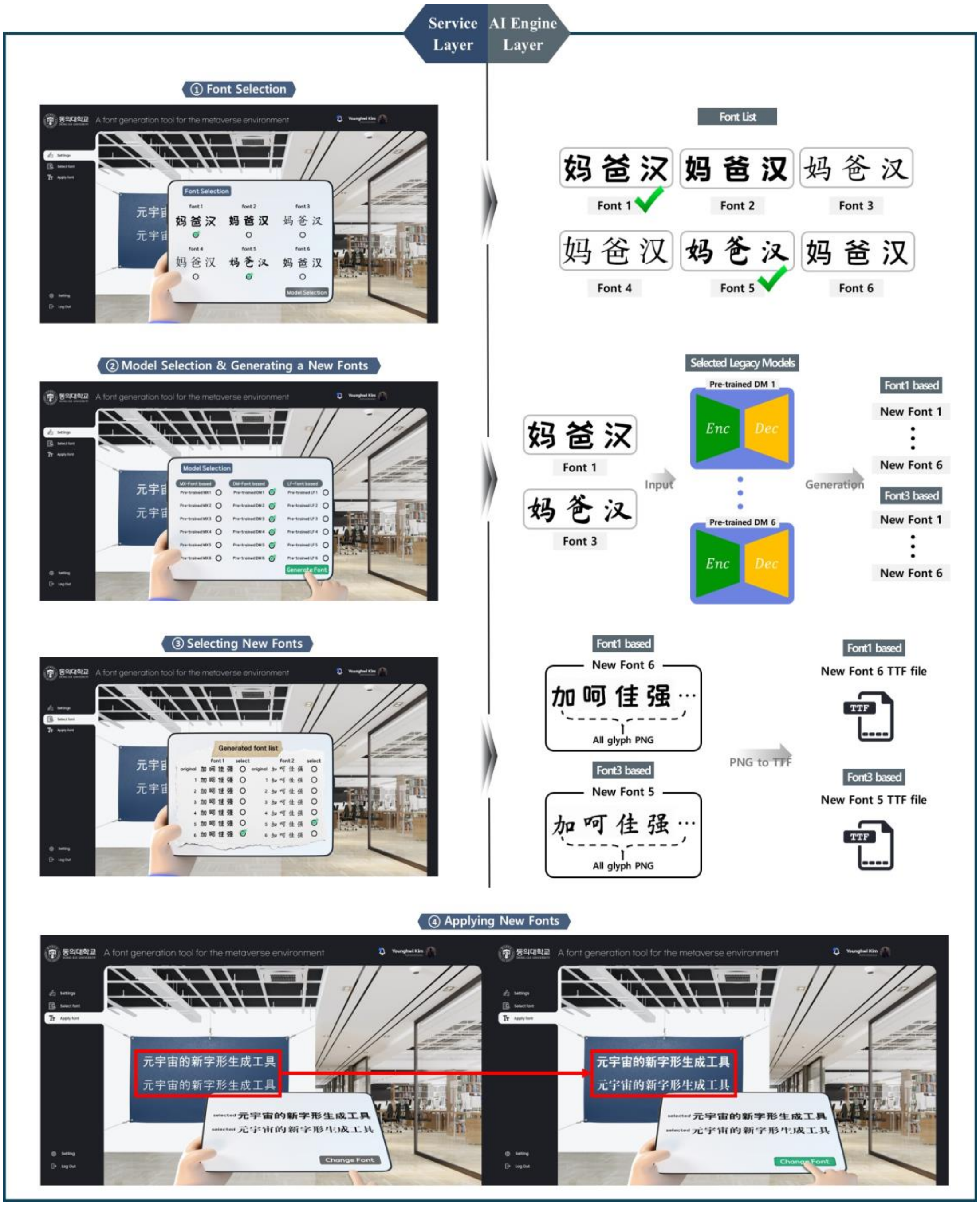

**Figure 5**. Overall framework of the automatic text design generation service for metaverse content.

The proposed service framework operates in a metaverse environment through the following steps: 1) Service users (such as metaverse content designers and metaverse users) select their preferred text design(s) in the metaverse development environment. 2) The system trains the model via legacy learning. 3) Newly designed text outputs are visualized for each trained model. 4) Once the desired text design is selected, it is saved as a True Type Font (TTF) file. 5) Thereafter, the saved TTF file is applied within the metaverse environment, enabling text design changes across all or selected contents. Figure 5 illustrates this framework.

## 5. Experiments

We conducted various performance evaluations to demonstrate the superiority of the proposed legacy learning and its text autogeneration service. To this end, we performed three experiments designed to answer the three research questions defined in the research objectives as follows.

- **(EX1: Performance of the legacy learning)** Can the proposed legacy learning method effectively generate new text designs regardless of the traditional FFG model approach?
- **(EX2: Qualitative and quantitative analysis of the effects of legacy learning)** Does repeated application of legacy learning lead to consistent transformations in text designs while maintaining high quality?
- **(EX3: User evaluation analysis of the automatic test design service)** Can the legacy learning-based automatic text design generation service serve as a practically useful tool for designers to create metaverse content?

### 5.1. Datasets

In the experiment, five Korean and five Chinese text design styles were used, considering factors such as stroke thickness, sharpness, and character spacing

**Table 2.** Visualization of text design samples for Chinese and Korean used in the experiments.

| Language | Notation | Text Design | Sample of Text Design |
|---|---|---|---|
| Chinese | TF1 | Droid Sans Fallback | 伽 倚 减 盖 团 裕 鹰 |
| | TF2 | FZ Kai Z03S | 伽 倚 减 盖 团 裕 鹰 |
| | TF3 | FZ Shu SongZ01S | 伽 倚 减 盖 团 裕 鹰 |
| | TF4 | Ma Shan Zheng | 伽 倚 减 盖 团 裕 鹰 |
| | TF5 | FZ Hei B01 | 伽 倚 减 盖 团 裕 鹰 |
| Korean | TF6 | Black Han Sans | 값 끗 돨 백 야 펴 힝 |
| | TF7 | GothicA1Medium | 값 끗 돨 백 야 펴 힝 |
| | TF8 | Hakgyoansim Santteutbatang | 값 끗 돨 백 야 펴 힝 |
| | TF9 | Nanum Brush | 값 끗 돨 백 야 펴 힝 |
| | TF10 | Noto Sans KR Extra | 값 끗 돨 백 야 펴 힝 |

Table 2 shows sample glyphs from the text designs used for training in each language. During the training process using the FFG model, each glyph was represented as an image with a resolution of 128 × 128 pixels and a single-channel image, and the batch size was set to 16. For the Korean dataset, a total of 2,500 glyph images were used. Among them, 80% (2,000 images) were used for training, while the remaining 20% (500 images) were held out for evaluation to assess the model's generalization ability to unseen characters. Similarly, for the Chinese dataset, 7,500 glyph images were used, with 80% (6,000 images) allocated for training and 20% (1,500 images) allocated for evaluation. The held-out glyphs were not included in the training data and were used exclusively to evaluate the generalization performance.

### 5.2. Baseline model and implementation details

***EX1 and EX2.*** For legacy learning, it is necessary to select a learning network structure from existing FFG models. In our experiments, legacy learning was applied to seven state-of-the-art high-performance FFG models: the DM-Font (Cha et al., 2020), LF-Font (Park et al., 2021a), MX-Font (Park et al., 2021b), CF-Font (Wang et al., 2023), VQ-Font (Yao et al., 2024), DIFF-Font (He et al., 2024c), and Font Diffuser (Yang et al., 2024) models. FFG models have been developed to generate text designs for specific languages. For example, DM-Font was designed to generate text designs exclusively for Korean datasets, whereas LF-Font, CF-Font, VQ-Font and Font Diffuser were designed to generate text designs for Chinese datasets. Consequently, legacy learning was applied to DM-Font to generate Korean text designs and to LF-Font, CF-Font, VQ-Font and Font Diffuser to generate Chinese text designs. By contrast, MX-Font and DIFF-Font can generate text designs for both Korean and Chinese; therefore, legacy learning was applied to both languages.

**Table 3**. Language compatibility of each few-shot font generation model.

| Language | DM-Font (2020) | LF-Font (2021) | MX-Font (2021) | CF-Font (2023) | VQ-Font (2024) | DIFF-Font (2024) | Font Diffuser (2024) |
|---|---|---|---|---|---|---|---|
| Chinese | | √ | √ | √ | √ | √ | √ |
| Korean | √ | | √ | | | √ | |

The training parameters were as follows: Legacy learning was conducted for six iterations with 200,000 training steps per iteration. The Adam optimizer was used to optimize the G, D, and CLS. All the experiments were conducted on a single NVIDIA GeForce RTX 4090 GPU.

### 5.3. Evaluation metrics

***EX1 and EX2.*** In this study, we adopted three evaluation metrics to assess the quality of the generated text designs comprehensively: 1) ***Fréchet Inception Distance (FID)***, a perceptual-level metric that measures the statistical similarity between image distributions; 2) ***Learned Perceptual Image Patch Similarity (LPIPS)***: a perceptual-level metric that measures visual similarity via feature maps extracted from a pretrained network; and 3) ***Optical Character Recognition (OCR) accuracy***, a functional metric that evaluates the legibility and recognizability of the generated glyphs. Whereas the FID and LPIPS provide quantitative assessments of visual similarity between the generated and original designs, the OCR accuracy directly measures text legibility and recognition performance. This complements visual similarity metrics by capturing structural distortions and error accumulation that may occur during iterative training.

Equation (10) presents the formula for the FID. Here, $\hat{x}$ represents the generated text design, and $x$ represents the original text design. The term $\mu$ denotes the mean extracted from the distribution of each text design, whereas $\sum$ represents the covariance matrix of the text design distribution. Therefore, the FID measures the similarity between two images and reflects the difference between their distributions.

$$FID(x,\hat{x}) = \|\mu_x - \mu_{\hat{x}}\|^2 + Tr\left(\sum_x + \sum_{\hat{x}} - \sqrt[2]{\sum_x \sum_{\hat{x}}}\right) \quad (10)$$

Equation (11) represents the formula for LPIPS: Here, $\hat{x}$ represents the generated text design, and $x$ represents the original text design. The term $l$ refers to the layer number of the pretrained VGG-16 network, $H$ and $W$ represent the height and width of the text design, respectively, and $h$ and $w$ are indices. Additionally, $w^l$ represents the weights extracted from layer $l$, and $x_{hw}^l$ and $\hat{x}_{hw}^l$ represent the feature maps at positions $h$ and $w$ from layer $l$, respectively. Thus, LPIPS measures the similarity between two text designs by summing the Euclidean distances between the feature maps across each layer of the text design.

$$LPIPS(x,\hat{x}) = \sum_l \frac{1}{H_l W_l} \sum_{h,w} \left\| w^l \odot (x_{hw}^l - \hat{x}_{hw}^l) \right\|_2^2). \quad (11)$$

The OCR accuracy evaluates the proportion of generated glyphs correctly recognized by an OCR engine. This metric directly assesses the legibility and recognizability of the generated designs, capturing structural damage or accumulated distortions that visual metrics may fail to detect. Following prior work (Cui et al., 2025), we computed the OCR accuracy by comparing the recognized text from the generated glyphs against the ground-truth labels.

***EX3.*** We performed a usability evaluation of the proposed legacy learning and automatic text design generation service for metaverse content on the basis of legacy learning. For this evaluation, we conducted a survey of 35 multidisciplinary designers with expertise in gaming, metaverse, and interface design from five IT companies, as well as 35 nondesigners. The survey consisted of two parts: 1) quantitative user evaluation and 2) SUS.

The survey questions for the quantitative user evaluation were structured from three perspectives: 1) generation perspective: *"How novel is the text design provided by the service compared with existing text designs?"*; 2) readability perspective: *"How good is the readability of the new text designs provided by the service?"*; and 3) design perspective: *"How satisfactory is the design of the new text designs provided by the service?"* Each perspective consists of ten questions, all of which are evaluated via a 5-point Likert scale. The survey was conducted as follows: Step 1) Respondents were presented with three text designs before the application of the automatic text design generation service. Step 2) Three newly generated text designs, created on the basis of the original designs via the automatic text design generation service, are presented. Step 3) The respondents compared the differences before and after completing the survey.

We used the SUS to evaluate the usability and practical applicability of the automatic text design generation service for metaverse content, which is based on legacy learning. The SUS is a widely used tool for evaluating the usability of various products and systems, such as websites and mobile phones (Tassabehji & Kamala, 2012). It provides the most reliable subjective evaluation results, regardless of sample size (Granić, Mitrović, & Marangunić, 2011). The SUS consists of ten questions related to usability, and respondents are given response options ranging from "Strongly Disagree (1)" to "Strongly Agree (5). The questions are divided into five odd-numbered positive items and five even-numbered negative items, with scores calculated to yield interpretable scores ranging from 0--100. Equation (12), where $U$ represents the item and $n$ represents the item number, is expressed as follows:

$$SUS\ score = 2.5 \times [\sum_{n=1}^{5}(U_{2n-1} - 1) + (5 - U_{2n})] \tag{12}$$

Figure 6 illustrates the criteria for evaluating usability levels based on the SUS scores, as expressed in Equation (12). Through this process, we can evaluate the usability and practical applicability of the proposed automatic text-design generation service for metaverse content.

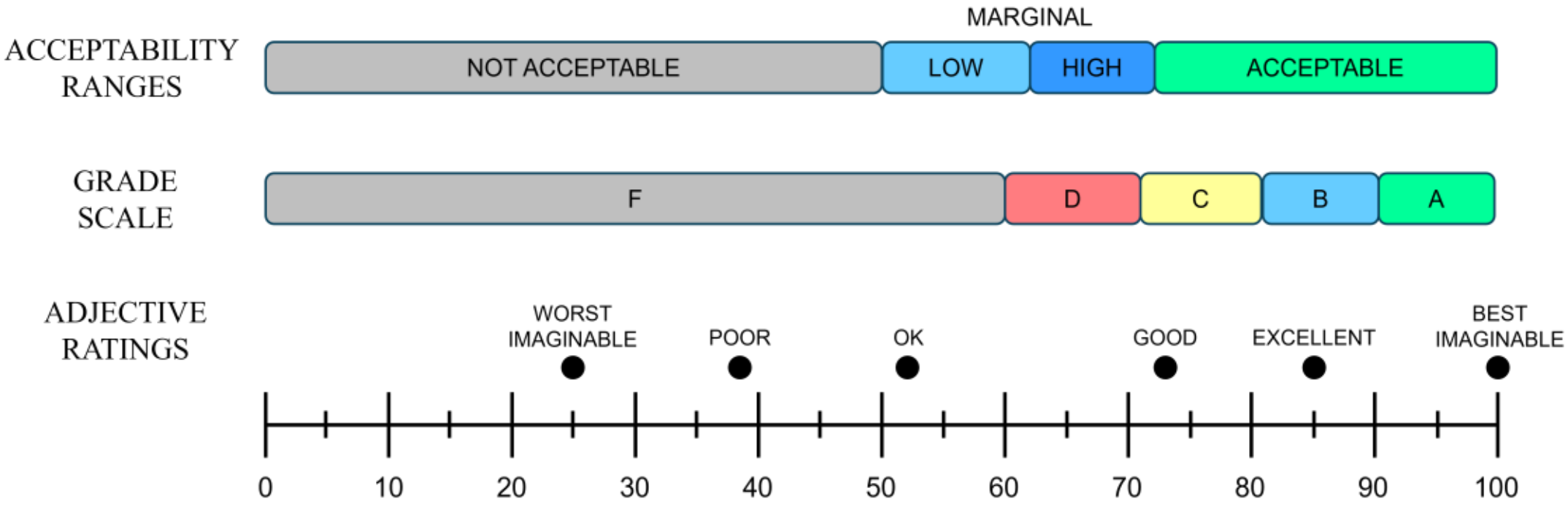

**Figure 6.** SUS acceptability ranges, grade scales, and adjective ratings.

## 6. Experimental results

### 6.1. Performance of Legacy Learning

Tables 4 to 7 present the experimental results comparing the effects before and after applying legacy learning, with legacy learning applied six times in all the experiments. Each table shows the LPIPS and FID results to measure the differences between the input text designs and the generated text designs.

Tables 4 and 5 compare different Chinese fonts. From the LPIPS perspective, there were differences between the FFG models for Chinese fonts; however, when legacy learning was applied, the text design changed by at least 12%, with a maximum change of up to 4.6 times. From the FID perspective, for Chinese fonts, when legacy learning was applied, the text design changed by at least 31%, with a maximum change of up to 4.2 times. Similarly, in the experiments for Korean text design, a greater change in text design occurred when legacy learning was applied than when the original text design was used. As shown in Table 6, in the LPIPS results, the text design changed by at least 37%, with a maximum change of up to six times. As shown in Table 7, in the FID results, the text design changed by at least two times, with a maximum change of up to seven times. As seen from these results, the legacy learning proposed in this study generates text designs that are transformed by up to 4.6 to 6 times from the original text design in terms of LPIPS and by up to 4.2 to 7 times in terms of FID, regardless of the model of the FFG or the text design used for training.

**Table 4**. LPIPS comparison of Chinese text designs before and after applying legacy learning across different FFG models.

| Model | | TF1 | TF2 | TF3 | TF4 | TF5 |
|---|---|---|---|---|---|---|
| Without Legacy Learning | LF-Font | 0.108 | 0.067 | 0.122 | 0.129 | 0.113 |
| | MX-Font | 0.156 | 0.132 | 0.194 | 0.165 | 0.183 |
| | CF-Font | 0.011 | 0.017 | 0.021 | 0.020 | 0.027 |
| | VQ-Font | 0.068 | 0.085 | 0.124 | 0.089 | 0.081 |
| | DIFF-Font | 0.164 | 0.166 | 0.184 | 0.194 | 0.181 |
| | Font Diffuser | 0.079 | 0.097 | 0.169 | 0.023 | 0.153 |
| With Legacy Learning | LF-Font | **0.121** | **0.081** | **0.134** | **0.132** | **0.133** |
| | MX-Font | **0.180** | **0.115** | **0.220** | **0.174** | **0.248** |
| | CF-Font | **0.022** | **0.064** | **0.054** | **0.049** | **0.283** |
| | VQ-Font | **0.120** | **0.115** | **0.167** | **0.131** | **0.119** |
| | DIFF-Font | **0.251** | **0.264** | **0.289** | **0.319** | **0.260** |
| | Font Diffuser | **0.329** | **0.624** | **0.212** | **0.319** | **0.369** |

**Table 5**. FID comparison of Chinese text designs before and after legacy learning is applied across different FFG models.

| Model | | TF1 | TF2 | TF3 | TF4 | TF5 |
|---|---|---|---|---|---|---|
| Without Legacy Learning | LF-Font | 81.26 | 45.19 | 58.72 | 71.61 | 45.98 |
| | MX-Font | 82.24 | 56.33 | 100.35 | 81.18 | 92.05 |
| | CF-Font | 9.48 | 32.98 | 19.61 | 17.31 | 40.63 |
| | VQ-Font | 88.48 | 73.84 | 108.27 | 76.57 | 97.32 |
| | DIFF-Font | 101.77 | 120.08 | 121.67 | 134.36 | 91.50 |
| | Font Diffuser | 36.46 | 26.91 | 25.74 | 58.84 | 39.54 |
| With Legacy Learning | LF-Font | **166.18** | **137.60** | **176.53** | **147.97** | **171.57** |
| | MX-Font | **175.41** | **87.06** | **206.18** | **137.37** | **232.54** |
| | CF-Font | **52.06** | **75.56** | **62.36** | **60.23** | **93.77** |
| | VQ-Font | **127.15** | **92.74** | **137.41** | **104.22** | **119.85** |
| | DIFF-Font | **133.14** | **146.96** | **178.23** | **186.84** | **118.66** |
| | Font Diffuser | **260.32** | **144.19** | **144.79** | **186.57** | **194.73** |

**Table 6.** LPIPS comparison of Korean text designs before and after legacy learning is applied across different FFG models.

| Model | | TF6 | TF7 | TF8 | TF9 | TF10 |
|---|---|---|---|---|---|---|
| Without Legacy Learning | DM-Font | 0.103 | 0.063 | 0.121 | 0.106 | 0.038 |
| | MX-Font | 0.152 | 0.112 | 0.110 | 0.159 | 0.092 |
| | DIFF-Font | 0.003 | 0.009 | 0.008 | 0.066 | 0.008 |
| With Legacy Learning | DM-Font | **0.139** | **0.100** | **0.142** | **0.136** | **0.057** |
| | MX-Font | **0.242** | **0.212** | **0.221** | **0.208** | **0.165** |
| | DIFF-Font | **0.027** | **0.090** | **0.031** | **0.289** | **0.086** |

**Table 7.** FID comparison of Korean text designs before and after legacy learning is applied across different FFG models.

| Model | | TF6 | TF7 | TF8 | TF9 | TF10 |
|---|---|---|---|---|---|---|
| Without | DM-Font | 97.44 | 48.47 | 53.72 | 68.94 | 44.30 |

| Legacy | MX-Font | 128.95 | 83.61 | 68.34 | 81.74 | 48.59 |
| --- | --- | --- | --- | --- | --- | --- |
| Learning | DIFF-Font | 8.26 | 12.47 | 22.28 | 14.63 | 10.41 |
| With | DM-Font | **218.41** | **203.02** | **189.73** | **179.19** | **141.64** |
| Legacy | MX-Font | **295.69** | **200.81** | **152.17** | **153.47** | **166.38** |
| Learning | DIFF-Font | **150.12** | **124.30** | **225.47** | **84.42** | **49.28** |

## 6.2. Analysis of the Legacy Learning effect

### 6.2.1. Quantitative analysis of the Legacy Learning effect

Figures 7 to 10 present the FID and LPIPS results for each model during the repeated application of legacy learning from the first to the sixth iteration, comparing the original text designs with the generated text designs.

According to the results, both the FID and LPIPS values consistently increased as legacy learning progressed. In the case of the Chinese text design, all the models exhibited upward-trending graphs, with LPIPS increasing by an average of 12.9% and FID increasing by an average of 23.2% per iteration. Similarly, in the Korean text design experiments, all the models displayed upward-trending graphs. On average, LPIPS increased by 22.0%, and FID increased by 35.2% per iteration.

Fundamentally, FFG models can learn the structural features of characters. However, owing to the limited diversity of the dataset, there may be constraints in perfectly replicating the reference style. On the basis of this characteristic, legacy learning induces gradual style variation across iterations while preserving the content structure. The extent and sensitivity of this style of transformation and its impact on quality can vary depending on the characteristics of the model.

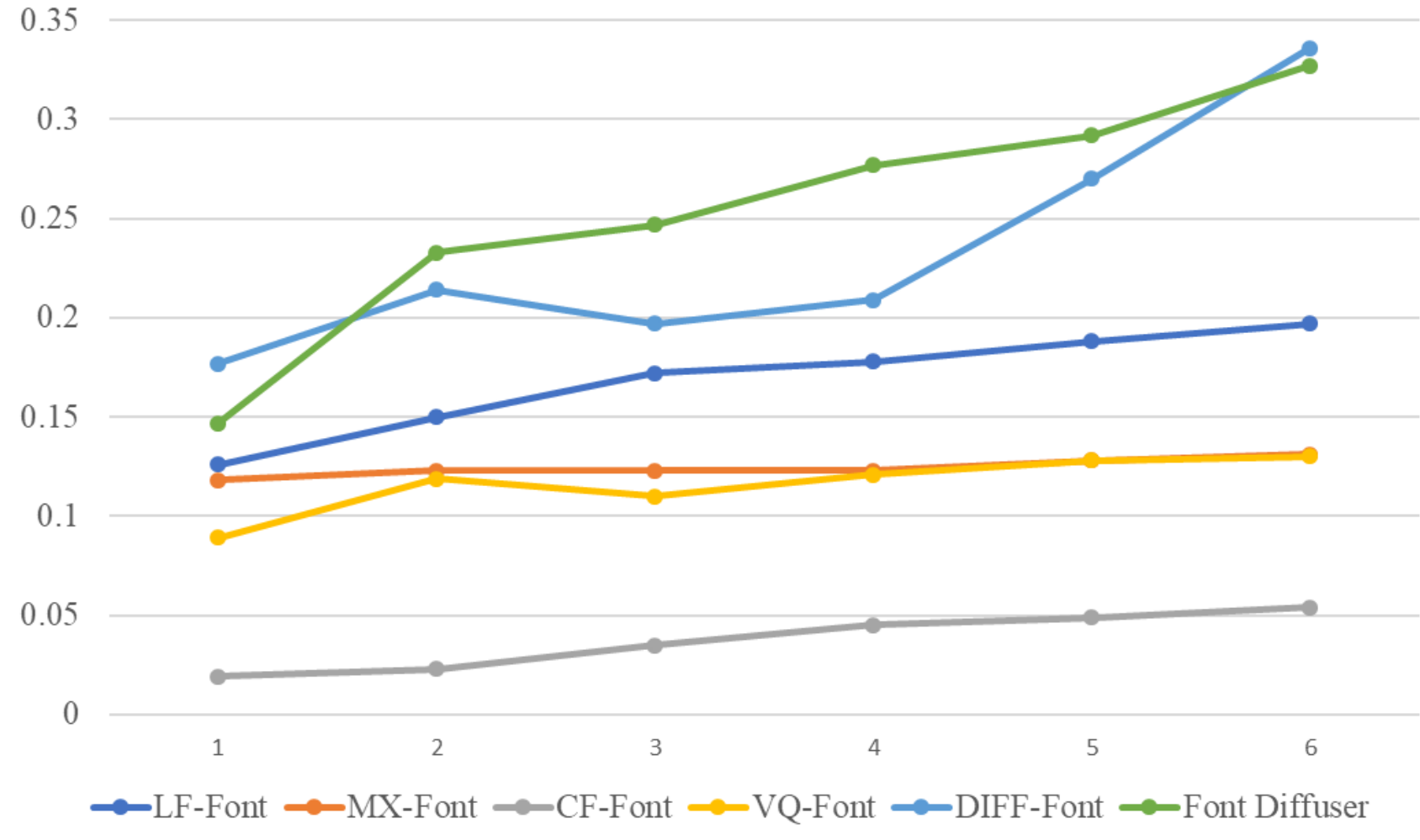

**Figure 7.** LPIPS trends over six legacy learning iterations for each model in the Chinese case.

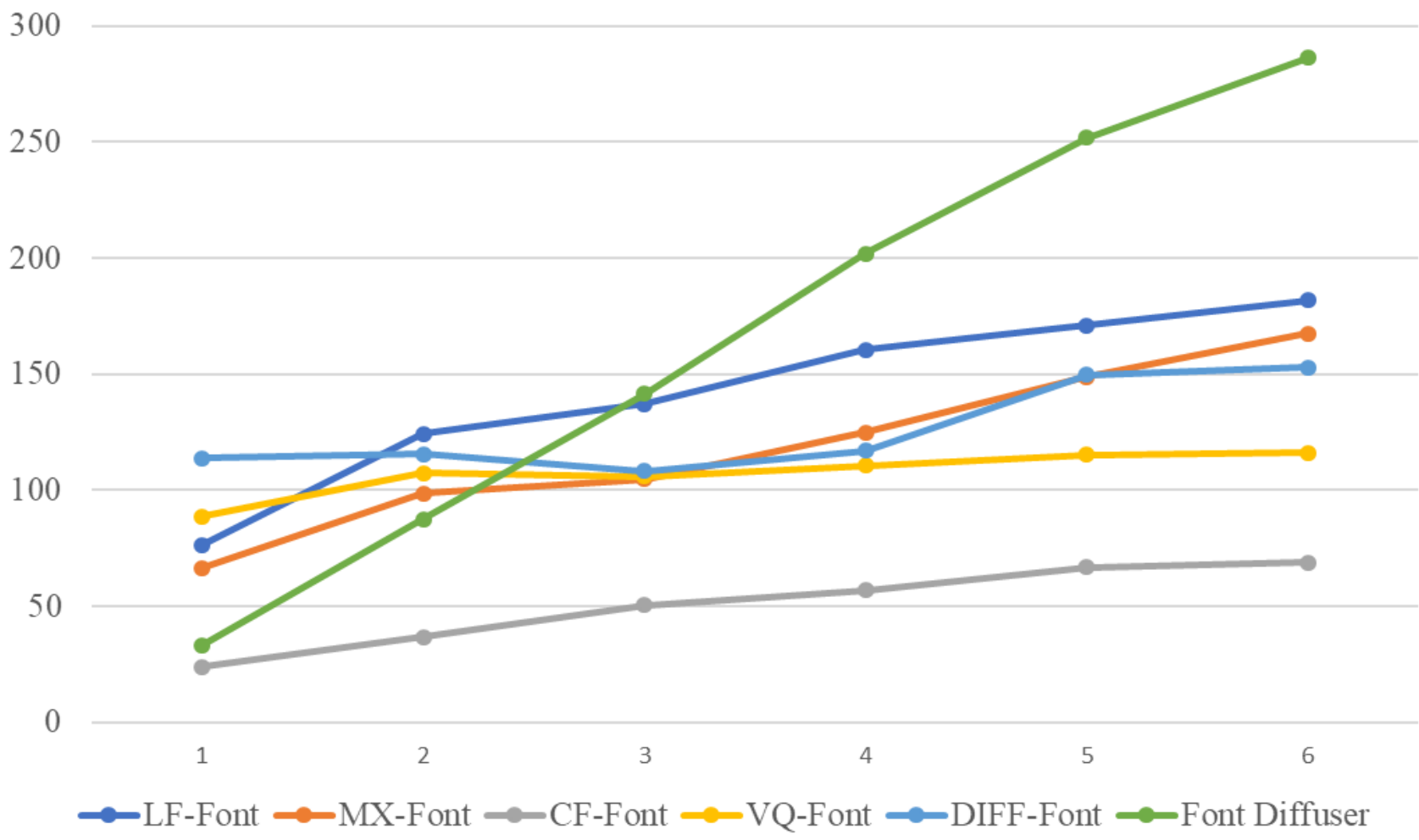

**Figure 8.** FID trends over six legacy learning iterations for each model in the Chinese case.

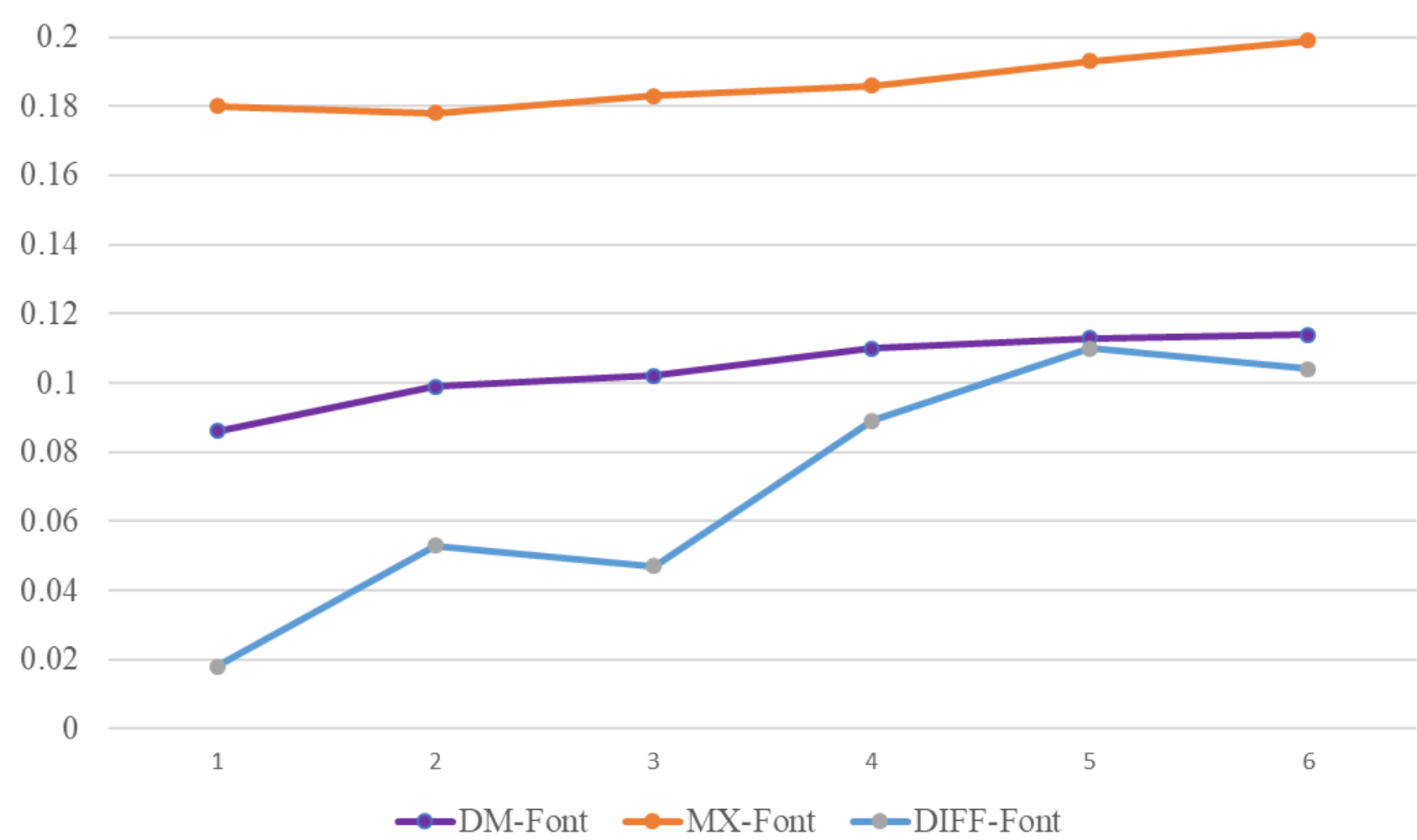

**Figure 9.** LPIPS trends over six legacy learning iterations for each model in the Korean case.

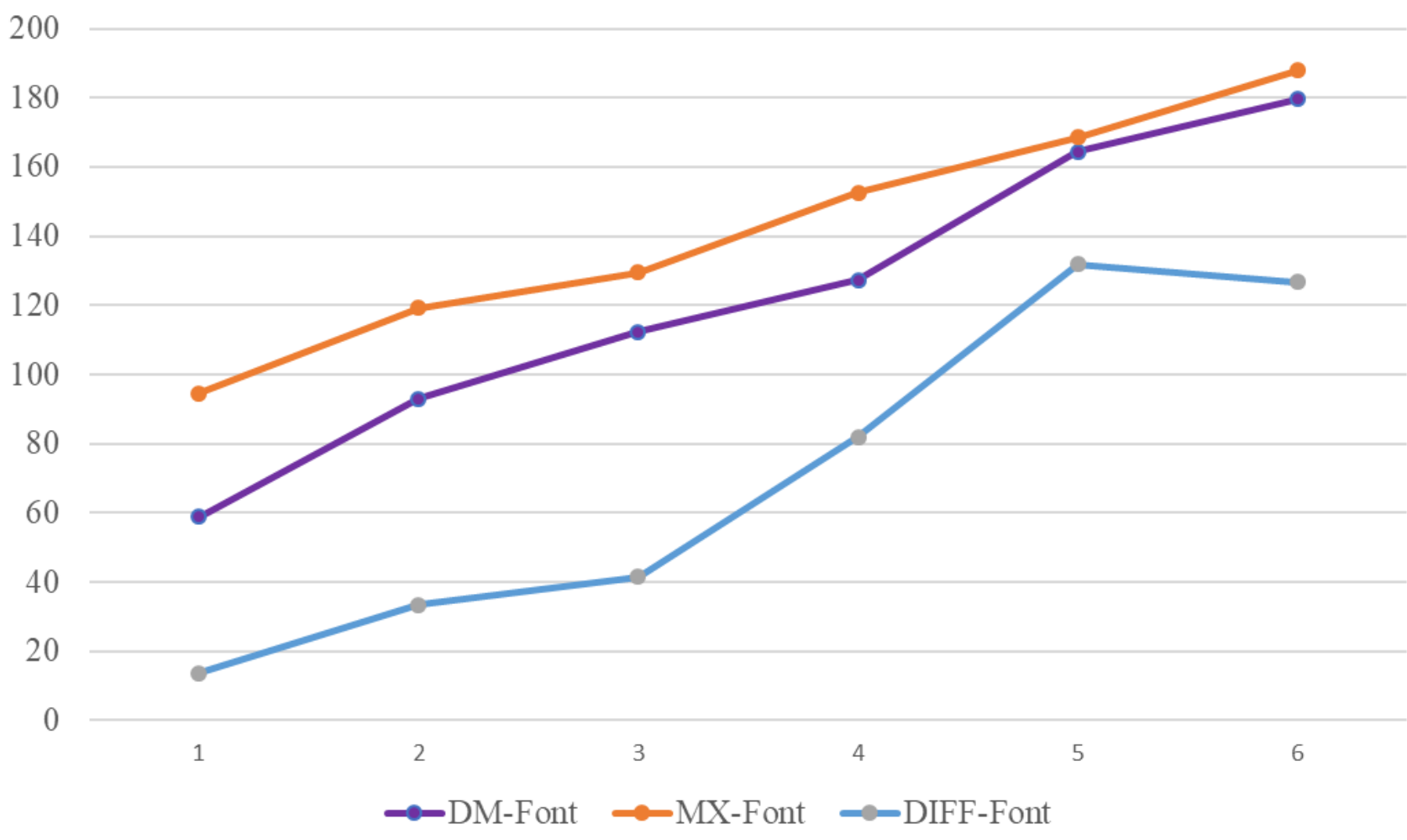

**Figure 10.** FID trends over six legacy learning iterations for each model in the Korean case.

In particular, CF-Font adopts an iterative style-vector refinement strategy to repeatedly refine the global style vector, aiming to produce minimally altered styles from a few input text designs. This structure enables the model to reproduce styles that closely resemble the original styles during the initial training, resulting in the smallest style changes even as legacy learning progresses. By contrast, Font Diffuser exhibits the opposite behavior. As a diffusion-based approach, the Font Diffuser initially shows small variation because of the goals of conventional FFG, but its multiscale fusion-based MCA adaptively integrates global and local information (Yang et al., 2024). Furthermore, its contrastive learning-based style disentanglement mechanism is highly sensitive to even subtle style shifts, resulting in the largest style variation overall. A similar trend was observed in the Korean text design experiments. MX-Font, which is designed to disentangle styles by components, consistently shows greater style variation than DM-Font. This finding indicates that an architecture that semantically localizes a global style representation can more readily accommodate diverse style transformations during the legacy learning process. These comparisons suggest that the effects of legacy learning iterations on style variation can differ depending on each model's inherent inductive bias and architectural design principles.

Figures 11 and 12 show the results of measuring the OCR accuracy for the Chinese and Korean text designs, respectively, at each iteration of legacy learning. The OCR accuracy was calculated by comparing the recognition results against the original text designs, and the similarity of the OCR to the original OCR per iteration was expressed as a percentage. Overall, the OCR results suggest that legacy learning not only transforms styles but also contributes to improvements in the functional recognizability of characters in most models. Although some misgenerated characters appeared in the early iterations of certain models, the overall trend showed a gradual increase in the OCR accuracy.

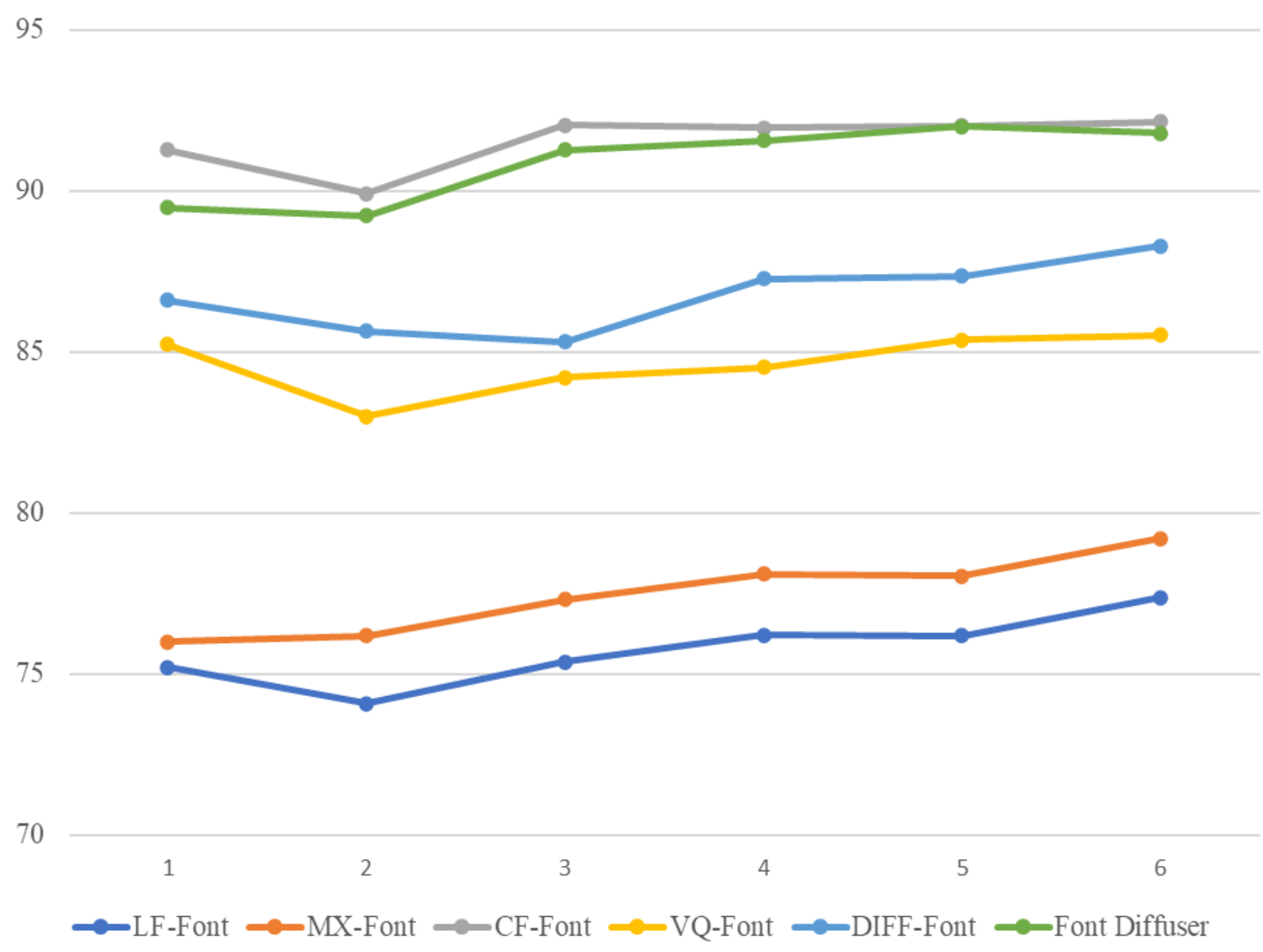

**Figure 11.** OCR accuracy variation over six legacy learning iterations in the Chinese case.

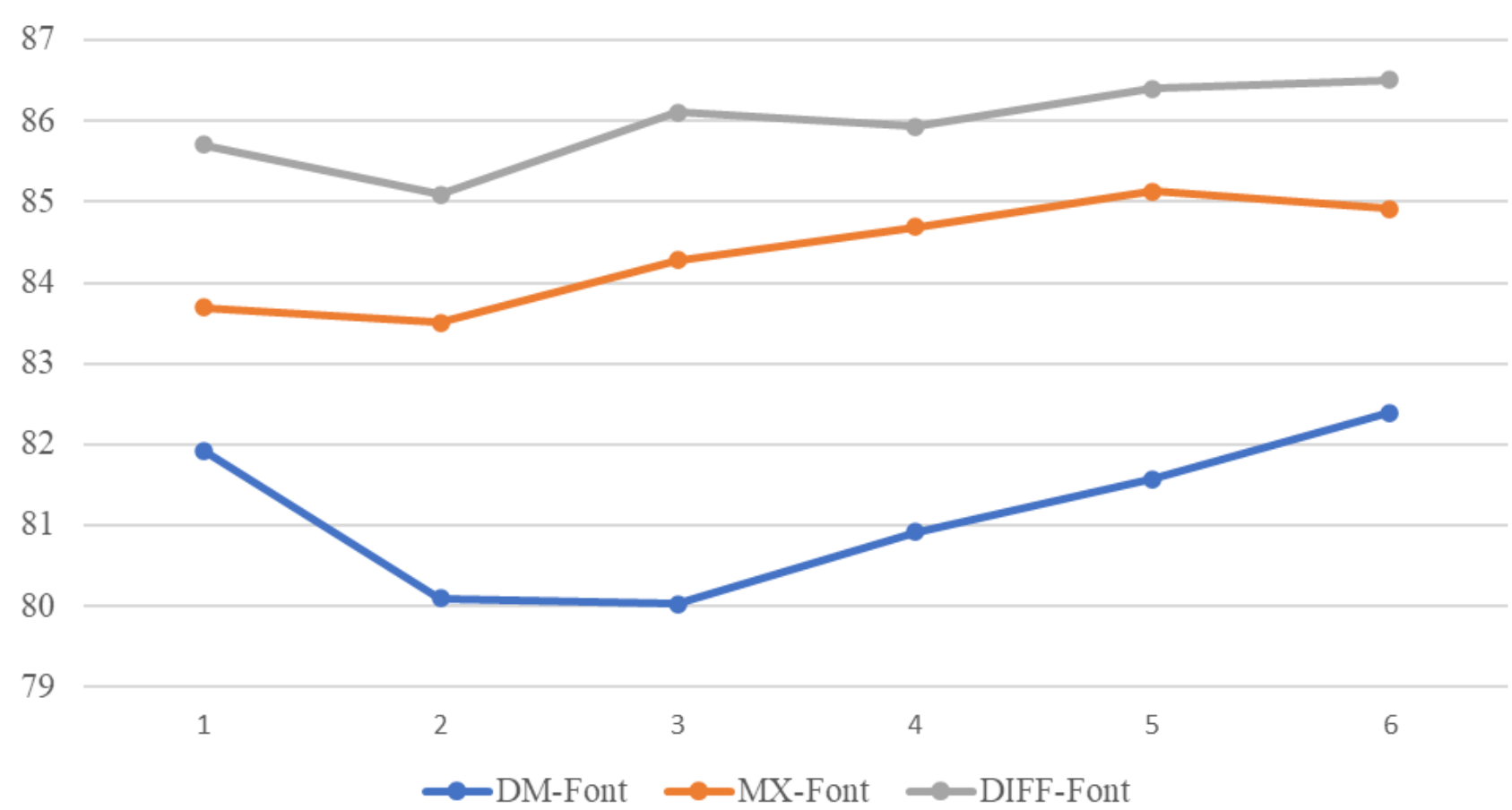

**Figure 12.** OCR accuracy variation over six legacy learning iterations in the Korean case

### 6.2.2. Qualitative analysis of the Legacy Learning effect

Figures 13 to 16 illustrate the visual quality of text designs generated by various text-generation models as legacy learning progresses. For visualization, fonts and characters were randomly selected from three models: DM-Font, MX-Font, and LF-Font. Samples were extracted following the application of legacy learning from the 1st to the 6th iteration.

As shown in the figures, in the early stages of legacy learning, the generated text designs resembled the original designs. However, in some cases, the generated glyphs contain noise, which degrades the overall quality. However, as legacy learning progressed, the designs gradually converged into more refined forms, and the initial noise tended to diminish.

These results suggest that while underfitting may occur in the initial stages, the repeated application of legacy learning gradually improves the quality of the text designs. In particular, after six iterations, the generated designs presented visual characteristics such as font size, stroke endings, and cursive styles that were distinct from those of the original designs.

**Figure 13.** Sample visualizations of Korean text designs generated through legacy learning via DM-Font.

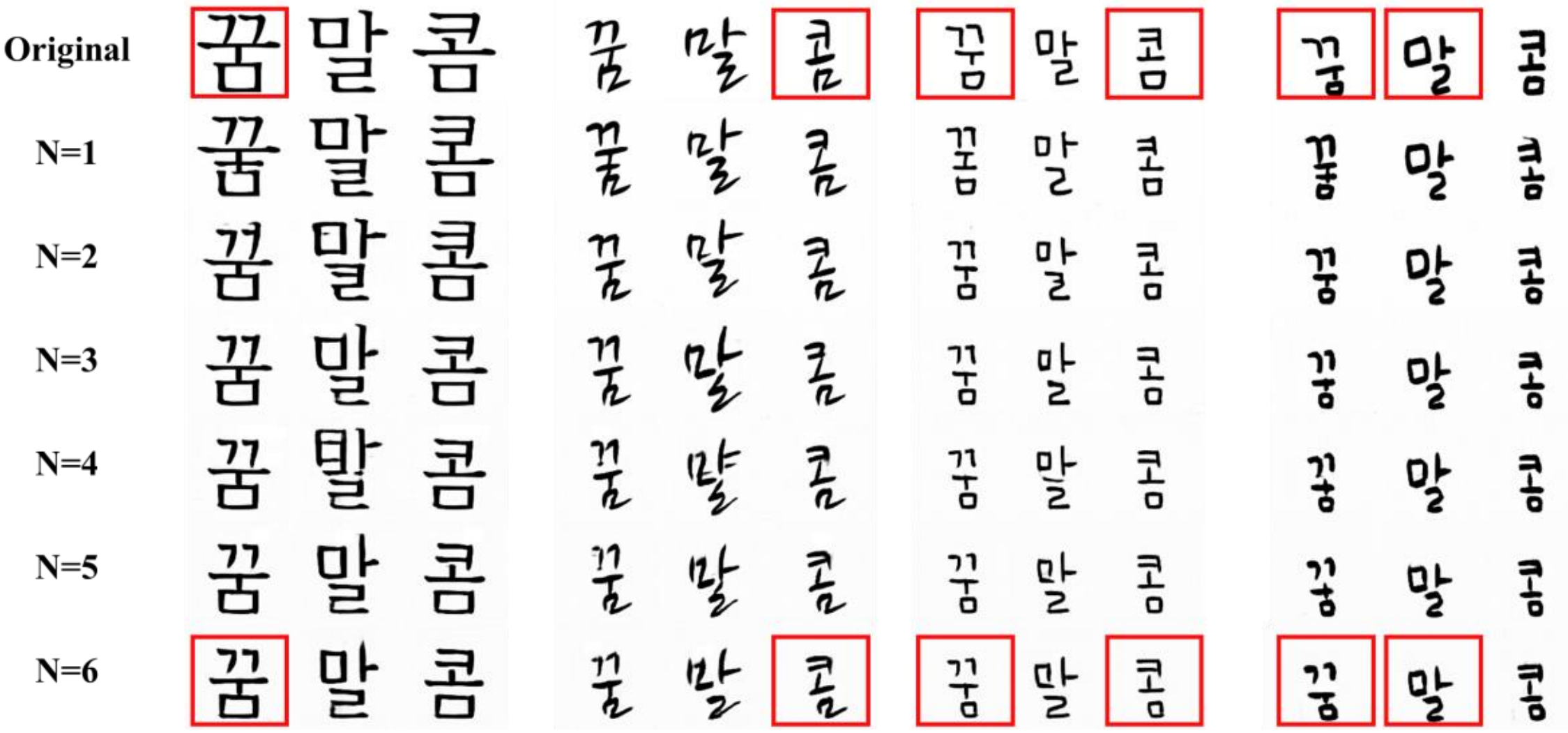

**Figure 14.** Sample visualizations of Korean text designs generated through legacy learning via MX-Font.

**Figure 15.** Sample visualizations of Chinese text designs generated through legacy learning via MX-Font.

**Figure 16.** Sample visualizations of Chinese text designs generated through legacy learning via LF-Font.

## 6.3. Practical applicability and user evaluation

### 6.3.1. Computational cost and failure case

To evaluate whether the proposed legacy learning strategy is practically applicable to metaverse environment construction, it is also necessary to consider the limitations in terms of computational resource requirements and system integration. This subsection discusses the computational cost of each model used in legacy learning and the failure cases observed during the generation process.

Table 8 summarizes the training time per iteration (200,000 steps), inference latency, and VRAM usage for each FFG model under identical settings. The VRAM requirements range from 9.6 to 18.3 GB (mean of 13.8 GB). A single legacy learning iteration takes 15.2 to 24.3 hours (mean 21.7 h). Inference is fast, 0.06 to 0.42 s per glyph, enabling on-demand rendering once a style is trained. These results suggest an offline training/online inference workflow: styles are trained and distributed in advance, whereas generation remains interactive. All the experiments were run on a consumer-grade NVIDIA GeForce RTX 4090 (<24 GB VRAM), indicating that both training and inference are feasible without data center class GPUs.

**Table 8.** VRAM usage, training time, and inference time of FFG models for Korean and Chinese glyph generation.

| | Model | VRAM usage (GB) | Training time per legacy iteration based on 200,000 training steps (h) | Inference time per glyph (s) |
|---|---|---|---|---|
| Korean | DM-Font | 14.8 | 22.6 | 0.28 |
| | MX-Font | 18.3 | 22.1 | 0.26 |
| | DIFF-Font | 12.2 | 23.2 | 0.38 |
| Chinese | LF-Font | 15.2 | 23.1 | 0.18 |
| | MX-Font | 18.3 | 23.5 | 0.16 |
| | CF-Font | 10.1 | 17.5 | 0.25 |
| | VQ-Font | 9.6 | 15.2 | 0.06 |
| | DIFF-Font | 12.2 | 24.3 | 0.32 |
| | Font Diffuser | 13.5 | 23.7 | 0.42 |

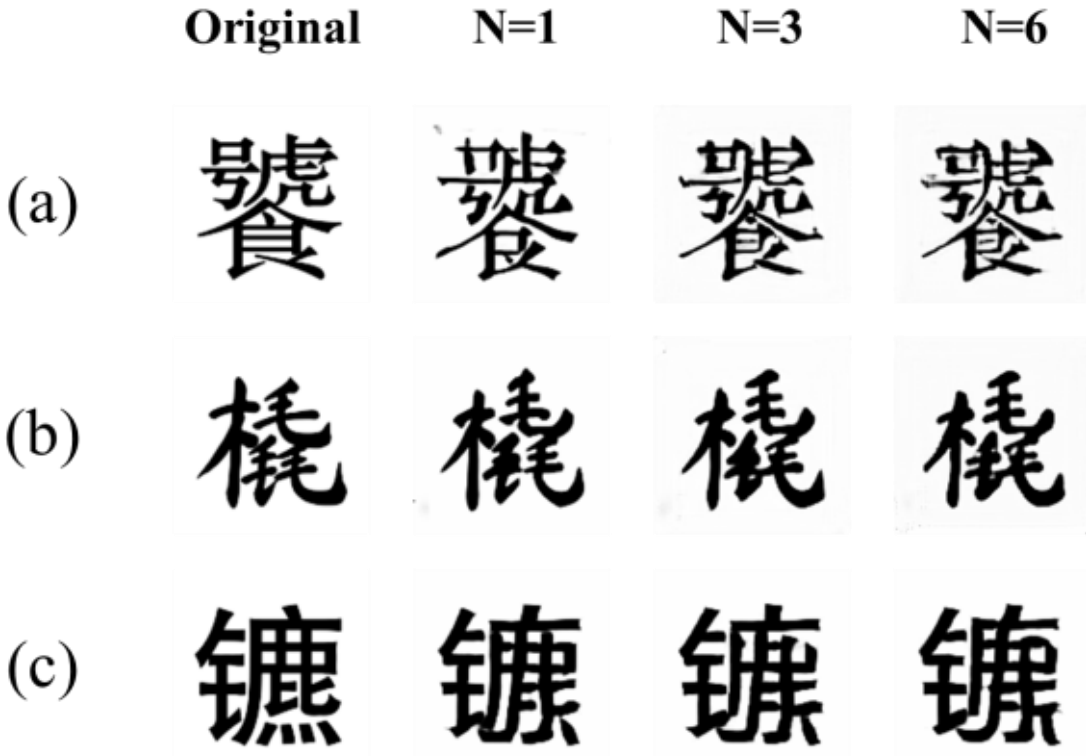

**Figure 17.** Failure cases: results for (a) MX-Font, (b) LF-Font, and (c) VQ-Font.

Figure 17 visualizes the failure cases observed during the Chinese glyph generation process of (a) MX-Font, (b) LF-Font, and (c) VQ-Font. In the early training stages, some generated glyphs exhibited errors, such as incomplete stroke representation or stroke segmentation. These issues are particularly prominent in characters with a large number of strokes or complex structures. Moreover, these problems were not resolved as legacy learning progressed. In some cases, the errors became even more pronounced through successive iterations.

### 6.3.2. Quantitative user evaluation.

Table 9 summarizes the statistics of the survey conducted by 35 multidisciplinary designers (Experts) and non-designers (Non-experts). According to the survey results, the average scores for the Generation, Readability, and Design perspectives were 4.53, 4.79, and 4.82, respectively, for the Expert group and 4.26, 4.85, and 4.68, respectively, for the Non-expert group. In particular, for the Expert group, the item "The generated text design of the transformed text design differs from the original text design" under the Generation perspective received a score of 4.91. From the Readability perspective, the item "The generated text design maintains readability" received a score of 5.00. From the Design perspective, the items "The generated text design is overall satisfying" and "The generated text design enhances design diversity" received scores of 4.91 and 4.94, respectively.

**Table 9.** Statistics of the survey conducted with both multidisciplinary designers and nondesigners are summarized, including the average and standard deviation for each.

| Group | Metric | Perspective | | |
|---|---|---|---|---|
| | | Generation | Readability | Design |
| Experts | Mean | 4.53 | 4.79 | 4.82 |
| | Standard Deviation | 0.09 | 0.07 | 0.08 |
| Nonexperts | Mean | 4.26 | 4.85 | 4.68 |
| | Standard Deviation | 0.33 | 0.12 | 0.45 |

### 6.3.3. System usability scale

Tables 10 and 11 summarize the results of the SUS survey conducted with 35 design experts and 35 non-experts. The values presented in the tables indicate the Likert scale responses for each item and the corresponding SUS scores calculated based on those responses.

In the expert group, the positive odd-numbered items generally received high scores ranging from 4 to 5 (average: 4.81), whereas the negative even-numbered items received low scores ranging from 1 to 2 (average: 1.15). By contrast, the non-expert group had relatively lower scores. The average scores for the odd- and even-numbered items were 3.88 and 1.76, respectively, which are 0.93 points lower and 0.61 points higher than those of the expert group. This discrepancy can be interpreted as stemming from differences in evaluation perspectives and criteria between the two groups. The expert group was more likely to analyze the quality, stylistic diversity, and applicability of the text designs from a technical or practical standpoint and thus may have provided positive assessments regarding the efficiency and applicability of the AI-based automated framework. By contrast, the non-expert group may have responded based on more intuitive and visual impressions, leading to relatively higher expectations for clear differentiation and completeness in the design.

**Table 10.** Summary of SUS scores and Likert-scale responses from design experts.

| User | Questions | | | | | | | | | | Score |
|---|---|---|---|---|---|---|---|---|---|---|---|
| | Q1 | Q2 | Q3 | Q4 | Q5 | Q6 | Q7 | Q8 | Q9 | Q10 | |
| 1 | 5 | 2 | 4 | 1 | 5 | 1 | 5 | 1 | 5 | 2 | 92.5 |
| 2 | 5 | 1 | 5 | 1 | 5 | 1 | 5 | 2 | 4 | 1 | 95.0 |
| 3 | 5 | 1 | 5 | 2 | 4 | 1 | 5 | 1 | 5 | 2 | 92.5 |
| 4 | 4 | 1 | 4 | 1 | 5 | 1 | 5 | 1 | 5 | 1 | 95.0 |
| 5 | 5 | 1 | 5 | 1 | 5 | 1 | 4 | 1 | 5 | 1 | 97.5 |
| 6 | 5 | 2 | 5 | 1 | 5 | 1 | 5 | 1 | 5 | 1 | 97.5 |
| 7 | 4 | 1 | 5 | 2 | 5 | 2 | 5 | 1 | 5 | 1 | 92.5 |
| 8 | 5 | 1 | 5 | 2 | 5 | 1 | 5 | 1 | 5 | 1 | 97.5 |
| 9 | 5 | 1 | 4 | 2 | 4 | 1 | 5 | 1 | 5 | 1 | 92.5 |
| 10 | 5 | 1 | 5 | 1 | 5 | 1 | 5 | 1 | 5 | 1 | 100 |
| 11 | 5 | 1 | 5 | 1 | 5 | 1 | 4 | 1 | 5 | 1 | 97.5 |
| 12 | 5 | 1 | 5 | 1 | 5 | 1 | 5 | 1 | 5 | 2 | 97.5 |
| 13 | 5 | 1 | 5 | 1 | 5 | 1 | 5 | 1 | 5 | 1 | 100.0 |
| 14 | 5 | 2 | 5 | 1 | 5 | 1 | 5 | 1 | 5 | 1 | 97.5 |
| 15 | 4 | 1 | 5 | 2 | 4 | 1 | 5 | 2 | 5 | 1 | 90.0 |
| 16 | 5 | 1 | 5 | 1 | 5 | 1 | 5 | 1 | 5 | 1 | 100 |
| 17 | 5 | 1 | 4 | 1 | 5 | 1 | 5 | 1 | 5 | 1 | 97.5 |
| 18 | 5 | 1 | 5 | 1 | 5 | 1 | 4 | 1 | 5 | 1 | 97.5 |
| 19 | 5 | 1 | 5 | 1 | 5 | 1 | 5 | 2 | 5 | 1 | 97.5 |
| 20 | 5 | 1 | 5 | 1 | 5 | 2 | 5 | 1 | 5 | 1 | 97.5 |
| 21 | 5 | 1 | 5 | 1 | 5 | 1 | 5 | 1 | 5 | 1 | 100.0 |
| 22 | 5 | 2 | 5 | 1 | 5 | 1 | 5 | 1 | 5 | 1 | 97.5 |
| 23 | 4 | 1 | 5 | 1 | 5 | 1 | 5 | 1 | 4 | 1 | 95.0 |
| 24 | 5 | 1 | 4 | 1 | 4 | 1 | 4 | 2 | 5 | 2 | 87.5 |
| 25 | 5 | 1 | 5 | 1 | 5 | 1 | 5 | 1 | 5 | 1 | 100.0 |
| 26 | 5 | 1 | 5 | 2 | 5 | 1 | 5 | 1 | 5 | 1 | 97.5 |
| 27 | 5 | 1 | 5 | 1 | 5 | 1 | 5 | 1 | 5 | 1 | 100 |
| 28 | 5 | 1 | 3 | 1 | 5 | 1 | 5 | 1 | 5 | 1 | 95.0 |
| 29 | 4 | 1 | 5 | 1 | 4 | 1 | 5 | 1 | 4 | 1 | 92.5 |
| 30 | 5 | 2 | 5 | 1 | 5 | 1 | 5 | 1 | 5 | 2 | 95.0 |
| 31 | 5 | 1 | 4 | 1 | 5 | 1 | 4 | 1 | 5 | 1 | 95.0 |
| 32 | 5 | 1 | 5 | 1 | 4 | 2 | 5 | 1 | 5 | 1 | 95.0 |
| 33 | 5 | 1 | 5 | 1 | 5 | 1 | 5 | 2 | 5 | 1 | 97.5 |
| 34 | 4 | 2 | 5 | 2 | 5 | 1 | 4 | 1 | 4 | 2 | 85.0 |
| 35 | 5 | 1 | 4 | 1 | 5 | 1 | 5 | 1 | 4 | 1 | 95.0 |
| Average | | | | | | | | | | | **95.78** |

In addition to these quantitative findings, the study applied interpretive standards such as "Acceptability ranges," "Grade scale," and "Adjective ratings," which have been widely used in previous studies, to provide a more nuanced qualitative evaluation. Accordingly, SUS scores were calculated via Equation (12), and the average score for each respondent group was obtained. As a result, the average SUS score for the 35 respondents in the expert group was 95.78, corresponding to “acceptable” in the Acceptability ranges, grade “A” in the Grade scale, and “Best Imaginable” in the Adjective ratings. For the non-expert group, the average SUS score was 76.42, corresponding to “acceptable,” grade “C,” and “Excellent,”, respectively.

**Table 11.** Summary of SUS scores and Likert-scale responses from non-experts.

| User | Questions | | | | | | | | | | Score |
|---|---|---|---|---|---|---|---|---|---|---|---|
| | Q1 | Q2 | Q3 | Q4 | Q5 | Q6 | Q7 | Q8 | Q9 | Q10 | |
| 1 | 4 | 3 | 4 | 2 | 3 | 1 | 3 | 3 | 4 | 2 | 67.5 |
| 2 | 3 | 3 | 5 | 1 | 3 | 1 | 4 | 2 | 4 | 1 | 77.5 |
| 3 | 4 | 1 | 3 | 1 | 4 | 2 | 3 | 2 | 4 | 1 | 77.5 |
| 4 | 4 | 1 | 3 | 3 | 5 | 3 | 4 | 1 | 4 | 2 | 75.0 |
| 5 | 3 | 2 | 3 | 3 | 3 | 2 | 4 | 2 | 4 | 1 | 67.5 |
| 6 | 5 | 2 | 5 | 3 | 4 | 3 | 3 | 1 | 3 | 1 | 75.0 |
| 7 | 4 | 1 | 5 | 1 | 3 | 2 | 3 | 3 | 3 | 3 | 70.0 |
| 8 | 4 | 1 | 3 | 3 | 4 | 2 | 4 | 1 | 4 | 1 | 77.5 |
| 9 | 5 | 2 | 4 | 1 | 4 | 1 | 4 | 2 | 5 | 2 | 85.0 |
| 10 | 3 | 2 | 4 | 2 | 4 | 2 | 4 | 2 | 3 | 1 | 72.5 |
| 11 | 3 | 2 | 4 | 2 | 5 | 1 | 4 | 2 | 3 | 1 | 77.5 |
| 12 | 3 | 1 | 5 | 3 | 4 | 3 | 5 | 1 | 5 | 2 | 80.0 |
| 13 | 5 | 3 | 4 | 1 | 3 | 2 | 5 | 3 | 3 | 2 | 72.5 |
| 14 | 4 | 1 | 3 | 1 | 5 | 1 | 4 | 2 | 4 | 3 | 80.0 |
| 15 | 5 | 1 | 4 | 1 | 3 | 1 | 4 | 2 | 3 | 1 | 82.5 |
| 16 | 5 | 1 | 3 | 2 | 5 | 1 | 4 | 2 | 3 | 2 | 80.0 |
| 17 | 4 | 2 | 4 | 2 | 3 | 3 | 4 | 1 | 3 | 2 | 70.0 |
| 18 | 4 | 1 | 4 | 1 | 4 | 1 | 4 | 1 | 5 | 1 | 90.0 |
| 19 | 4 | 1 | 5 | 3 | 3 | 2 | 5 | 2 | 4 | 1 | 80.0 |
| 20 | 3 | 3 | 5 | 2 | 4 | 1 | 4 | 3 | 5 | 2 | 75.0 |
| 21 | 4 | 1 | 3 | 3 | 4 | 1 | 3 | 2 | 3 | 1 | 72.5 |
| 22 | 4 | 2 | 4 | 1 | 4 | 1 | 4 | 2 | 4 | 3 | 77.5 |
| 23 | 4 | 2 | 3 | 1 | 3 | 1 | 5 | 1 | 5 | 2 | 82.5 |
| 24 | 4 | 2 | 4 | 2 | 3 | 2 | 3 | 1 | 4 | 1 | 75.0 |
| 25 | 5 | 3 | 5 | 1 | 3 | 3 | 4 | 3 | 5 | 3 | 72.5 |
| 26 | 3 | 1 | 5 | 1 | 3 | 2 | 5 | 2 | 4 | 1 | 82.5 |
| 27 | 4 | 3 | 3 | 2 | 5 | 2 | 4 | 2 | 4 | 1 | 75.0 |
| 28 | 5 | 3 | 5 | 3 | 4 | 2 | 4 | 2 | 4 | 2 | 75.0 |
| 29 | 5 | 1 | 4 | 1 | 4 | 1 | 4 | 2 | 3 | 2 | 82.5 |
| 30 | 3 | 1 | 5 | 1 | 3 | 2 | 4 | 1 | 4 | 1 | 82.5 |
| 31 | 4 | 3 | 4 | 2 | 3 | 1 | 3 | 3 | 4 | 2 | 67.5 |
| 32 | 3 | 3 | 5 | 1 | 3 | 1 | 4 | 2 | 4 | 1 | 77.5 |
| 33 | 4 | 1 | 3 | 1 | 4 | 2 | 3 | 2 | 4 | 1 | 77.5 |
| 34 | 4 | 1 | 3 | 3 | 5 | 3 | 4 | 1 | 4 | 2 | 75.0 |
| 35 | 3 | 2 | 3 | 3 | 3 | 2 | 4 | 2 | 4 | 1 | 67.5 |
| Average | | | | | | | | | | | **76.42** |

Although the non-expert group provided relatively lower evaluations, this suggests the need for improvements in enhancing the distinguishability of diverse text designs. By contrast, the expert group’s highly favorable assessments indicate that the proposed legacy learning-based service framework provides meaningful contributions in terms of its fully automated structure, data augmentation potential, and enhancement of existing FFG model performance. Considering the results of both the survey and the SUS evaluation, the proposed framework demonstrates high practical utility and suitability as an effective tool in metaverse environments.

## 7. Implications

### 7.1. Theoretical implications

Legacy learning expands the theoretical framework of text design by demonstrating the viability of iterative learning in generating high-quality and diverse outputs. This method challenges traditional approaches that rely on partial datasets and instead provides a comprehensive strategy for text design. The integration of entire datasets into the learning process enhances the consistency of generated text designs while also broadening the scope of application to include complex and multilingual text design tasks. This study provides a robust foundation for advancing AI-based text-generation methodologies in the metaverse and beyond. Furthermore, the introduction of style transformations within legacy learning represents a significant advancement toward understanding how iterative processes can drive innovation in text design. By systematically refining text designs through multiple iterations, this method underscores the importance of adaptability and refinement in generative AI applications. These advancements contribute to the theoretical discourse surrounding AI-driven design and its role in shaping digital user experiences.

### 7.2. Practical implications

From a practical perspective, legacy learning offers transformative potential for designers working within metaverse environments. By addressing the challenges associated with languages requiring extensive character sets, such as Korean and Chinese, legacy learning enables the efficient generation of high-quality text designs while minimizing manual input. This efficiency not only reduces costs but also significantly accelerates production timelines, making it an invaluable tool for large-scale metaverse content-creation projects. The usability testing results further highlight the practical utility of legacy learning. With an SUS score of 95.8, designers found the tool both intuitive and highly effective, particularly in streamlining workflows and enhancing creativity. These findings indicate that legacy learning not only meets the technical requirements of metaverse content design but also aligns with the operational needs of professional designers. Its seamless integration into existing workflows further demonstrates its potential to become a standard tool for text design in digital environments.

### 7.3. Limitations and future directions

The proposed legacy learning strategy has significant potential in enhancing the diversity, quality, and scalability of AI-based text design in metaverse environments. However, several limitations remain that should be acknowledged and addressed in future research.

First, the experiments conducted in this study are limited to Chinese and Korean, both of which possess relatively well-structured stroke-based systems and standardized character layouts. However, the generalizability of the proposed method to broader linguistic contexts has yet to be verified. For example, languages such as Arabic, which feature cursive structures, and Vietnamese, which includes complex diacritics, present fundamentally different technical challenges. These scripts often demand context-aware stroke transitions or advanced spatial alignment. Future research should explore the adaptability of the proposed method to linguistically and structurally diverse scripts and investigate the application of language-specific refinement modules after each legacy learning iteration to increase performance.

Second, the current model naturally induces stylistic variation through data composition and iterative training, but it does not explicitly control the direction or intensity of design transformation. This may limit the applicability of the framework in user-oriented or style-guided text design tasks. Whereas this study focuses on the potential of automated style diversification, future studies should aim to systematically categorize and quantitatively evaluate the direction and degree of design variation according to the number of input glyphs and the types of design combinations, thereby providing insights into the controllability of style transformation within legacy learning.

Third, in the case of complex characters, especially those with multilayered stroke structures, error accumulation during the iterative training process was observed. Although legacy learning contributes to the enhancement of visual diversity, it may also amplify minor initial errors, leading to a decrease in structural accuracy in highly complex glyphs. To mitigate these limitations, refinement mechanisms before and after each iteration need to be incorporated. Future research could explore incorporating structural preservation mechanisms, such as selective memory modules or iteration-specific error correction pathways.

## 8. Conclusions

This study proposes a novel AI-based text design generation method known as legacy learning, which is specifically tailored to create text designs within metaverse content. We also introduce a service framework that combines this learning method with automatic text-design generation in metaverse environments. Text design for metaverse content is crucial for enhancing user readability, improving community engagement, and increasing immersion.

Text design tasks vary significantly according to language. For example, languages such as Korean and Chinese require the design of tens of thousands of characters, which poses significant challenges. The results of this study offer a new approach to efficiently generate and modify text designs for metaverse content at a lower cost.

Recently, FFG models using generative AI have been developed for text design. However, these models frequently produce text designs with noise, limiting their applicability to automatic generation and metaverse content. By contrast, the proposed legacy learning method overcomes the limitations of existing FFG models by enabling the creation of new high-quality text designs suitable for direct use and application in metaverse environments. Quantitative and qualitative evaluations were conducted to validate the text designs generated through legacy learning. The quantitative evaluation using the FID and LPIPS metrics revealed differences of 17.45% and 29.2%, respectively, per legacy learning iteration. In addition, the OCR-based evaluation confirmed that the readability of the generated text designs improved progressively with each iteration. The qualitative evaluation also confirmed that the generated text designs maintained high quality while presenting new stylistic variations based on the original designs. For the service framework applying legacy learning to automatic text design in metaverse content, usability testing was conducted with both designer and non-designer user groups. The resulting SUS scores were 95.8 and 76.4, respectively, demonstrating the practical utility and applicability of the proposed service in real-world metaverse content-creation scenarios.

The legacy learning method and service framework developed in this study have significant potential as valuable tools for text design in metaverse environments. This study is expected to stimulate further studies on AI-based services for metaverse creation.